\documentclass[aps,pre,twocolumn,amsmath,amssymb,floatfix,footinbib,superscriptaddress,letterpaper]{revtex4-1}

\let\SavedSigma\Sigma

\usepackage{amsmath,amssymb,mathtools}
\usepackage{stmaryrd}
\usepackage{times}
\usepackage[euler]{textgreek}
\usepackage[bbgreekl]{mathbbol}
\usepackage{graphicx}
\usepackage{bm,color}
\usepackage{natbib,hyperref}
\usepackage[capitalize]{cleveref}

\hypersetup{colorlinks=true,linkcolor=blue,citecolor=blue,urlcolor=blue}

\renewcommand{\rho}{\text{\textrho}}
\renewcommand{\sigma}{\text{\textsigma}}
\newcommand{\oket}[1]{\left|#1\right)}
\newcommand{\oketb}[1]{\Big|#1\Big)}
\newcommand{\oketn}[1]{|#1)}

\newcommand{\omatb}[3]{\Big( #1 \Big| #2 \Big| #3 \Big)}

\newcommand{\be}{\begin{equation}}
\newcommand{\ee}{\end{equation}}
\newcommand{\nn}{\nonumber}
\newcommand{\lf}{\left}
\newcommand{\rt}{\right}

\DeclareMathOperator{\Tr}{tr}

\newcommand{\dg}{^{\da}}

\newcommand{\da}{\dagger}
\newcommand{\dda}{\da \hspace{-2pt} \da}

\newcommand{\ginv}{^{-1}}

\newcommand{\alp}{\alpha}
\newcommand{\ep}{\epsilon}
\newcommand{\ld}{\lambda}
\newcommand{\s}{\sigma}

\newcommand{\dw}{\delta \omega}

\newcommand{\dwrm}[1]{\dw_{#1}}
\newcommand{\dwrk}{\dwrm{\vk}}
\newcommand{\dwq}{\delta \Omega}

\newcommand{\ga}{\gamma}
\newcommand{\gar}{\ga}
\newcommand{\gaq}{\Gamma}

\newcommand{\epe}{\ep_{\mathrm{q}}}

\newcommand{\at}{\tilde{\mathsf{a}}}
\newcommand{\bt}{\mathsf{b}}
\newcommand{\sfH}{\mathsf{H}}

\newcommand{\vm}{\mathbf{0}}
\newcommand{\vk}{\mathbf{k}}
\newcommand{\vp}{\mathbf{r}}

\newcommand{\ka}{\kappa}

\newcommand{\abs}[1]{\lf| #1 \rt|} 

\newcommand{\sTrb}[1]{\mathrm{tr} [ #1 ]}

\newcommand{\savg}[1]{\langle #1 \rangle}

\newcommand{\inner}[2]{\left(#1|#2\right)}
\newcommand{\sinner}[2]{( #1|#2 )}

\newcommand{\sCom}[1]{\lf[#1 , {}_\bullet \rt]}

\newcommand{\si}[2]{\s^{#1}_{#2}}
\newcommand{\spl}[1]{\si{+}{#1}}
\newcommand{\smi}[1]{\si{-}{#1}}

\newcommand{\spvp}{\spl{\vp}}
\newcommand{\smvp}{\smi{\vp}}

\newcommand{\sNmaev}[1]{t_{#1}}

\newcommand{\sbasym}{\mathbb{b}}
\newcommand{\sbbsym}{\bbbeta}
\newcommand{\sba}[1]{\sbasym_{#1}}
\newcommand{\sbb}[1]{\sbbsym_{#1}}
\newcommand{\sbap}[1]{\sbasym^{\dda}_{#1}}
\newcommand{\sbbp}[1]{\sbbsym^{\dda}_{#1}}

\newcommand{\sL}{\mathbb{L}}
\newcommand{\sLr}[1]{\sL_{r}^{#1}}
\newcommand{\sLq}[1]{\sL_{q}^{#1}}

\newcommand{\sA}{\mathbb{A}}
\newcommand{\sD}{\mathbb{D}}

\newcommand{\spf}[1]{\mathbb{f}_{#1}}
\newcommand{\spSE}[1]{\bbsigma_{#1}}

\newcommand{\sU}{\mathbb{U}}
\newcommand{\sUp}{\mathbb{T}}
\newcommand{\sUpc}[1]{\sUp_{#1}}

\newcommand{\sUkp}[2]{\sU^{#1 #2}}

\newcommand{\selfe}{\mathbb{\SavedSigma}}
\newcommand{\sloop}{\mathbb{\SavedSigma}}

\newcommand{\sloopn}[1]{\sloop_{#1}}
\newcommand{\evasloop}{\Sigma}

\newcommand{\evr}[1]{\mathsf{u}_{#1}}
\newcommand{\evl}[1]{\breve{\mathsf{u}}_{#1}}
\newcommand{\eva}[1]{\mathsf{\ld}_{#1}}

\newcommand{\evrc}[2]{\mathsf{u}_{#1}^{#2}}
\newcommand{\evlc}[2]{\breve{\mathsf{u}}_{#1}^{#2}}
\newcommand{\evac}[2]{\ld_{#1}^{#2}}

\newcommand{\evrr}[3]{\mathsf{r}^{#1}_{#2#3}}
\newcommand{\evlr}[3]{\breve{\mathsf{r}}^{#1}_{#2#3}}
\newcommand{\evar}[3]{\mathsf{\ld}^{#1}_{#2#3}}
\newcommand{\evrrvk}[2]{\evrr{\vk}{#1}{#2}}
\newcommand{\evarvk}[2]{\evar{\vk}{#1}{#2}}

\newcommand{\evrs}[1]{\evrr{#1}{0}{0}}

\newcommand{\evrq}[2]{\mathsf{q}^{#1}_{#2}}
\newcommand{\evlq}[2]{\breve{\mathsf{q}}^{#1}_{#2}}
\newcommand{\evaq}[2]{\ell^{#1}_{#2}}
\newcommand{\evrqvp}[1]{\evrq{\vp}{#1}}
\newcommand{\evlqvp}[1]{\evlq{\vp}{#1}}
\newcommand{\evaqvp}[1]{\evaq{\vp}{#1}}
\newcommand{\evqs}[1]{\evrq{#1}{0}}

\newcommand{\rs}{\rho_\mathrm{s}}
\newcommand{\rsc}[1]{\rho^{[ #1]}}
\newcommand{\rscp}[2]{\rho_{#2}^{[ #1]}}
\newcommand{\rscl}[1]{\breve{\rho}^{[ #1]}}
\newcommand{\rscpl}[2]{\breve{\rho}_{#2}^{[ #1]}}

\newcommand{\rsoc}[1]{\rho_{ #1}}

\newcommand{\msar}{\mathsf{a}_{\vp}}

\begin{document}
\title{Resummation for Nonequilibrium Perturbation Theory and Application to Open Quantum Lattices}
\author{Andy C.~Y.~Li}
\affiliation{Department of Physics and Astronomy, Northwestern University, Evanston, Illinois 60208, USA}
\author{F. Petruccione}
\affiliation{Quantum Research Group, School of Chemistry and Physics, University of KwaZulu-Natal, Durban 4001, South Africa, and
National Institute for Theoretical Physics (NITheP), KwaZulu-Natal, South Africa}
\author{Jens Koch}
\affiliation{Department of Physics and Astronomy, Northwestern University, Evanston, Illinois 60208, USA}
\date{\today}

\begin{abstract}
	Lattice models of fermions, bosons, and spins have long served to elucidate the essential physics of quantum phase transitions in a variety of systems. Generalizing  such models to incorporate driving and dissipation has opened new vistas to investigate nonequilibrium phenomena and dissipative phase transitions in interacting many-body systems. 
	We present a framework for the treatment of such open quantum lattices based on a resummation scheme for the Lindblad perturbation series. Employing a convenient diagrammatic representation, we utilize this method to obtain relevant observables for the open Jaynes-Cummings lattice, a  model of special interest for open-system quantum simulation. We demonstrate that the resummation framework allows us to reliably predict observables for both finite and infinite Jaynes-Cummings lattices with different lattice geometries. The resummation of the Lindblad perturbation series can thus serve as a valuable tool in validating open quantum simulators, such as circuit-QED lattices, currently being investigated experimentally. 
\end{abstract}

\maketitle
\section{Introduction}
\label{sec:introduction}
Lattice models describe particles or spins residing on a set of sites, arranged in a regular fashion. Different types of interactions among these components are possible and can be included in the formulation of the model. For this reason, lattice models can cover a large arena of physical systems and phenomena. Prominent examples are the fermionic Hubbard model \cite{Gutzwiller1963,Hubbard1963,Hubbard1964}, the Bose-Hubbard model \cite{Fisher1989,Krauth1992}, and the Heisenberg model \cite{Orbach1958,Anderson1973}.

\emph{Open} quantum lattices extend the lattice-model concept. They  include the effects of an environment and external driving fields coupled to the lattice. Such open lattices generally describe nonequilibrium phenomena and are of great interest in many subfields of physics -- ranging from condensed matter \cite{Takei2008,Sedlmayr2013,Lee2014} and AMO physics \cite{Carusotto2009,Ozawa2014,Hafezi2014,Raftery2014,Kapit2014,Peano2015,Gao2015} to applications in quantum information \cite{Devoret2013a,Boixo2014,Schuld2014,Pavlichin2014,Nickerson2014,Aron2014}.
Recently, many studies studying open quantum lattices have advanced our understanding of many-body systems under nonequilibrium conditions \cite{Baumann2010,Raftery2014,Gao2015}. Examples of phenomena predicted to emerge in certain scenarios include nonequilibrium critical behavior \cite{Takei2008,Carusotto2009,Baumann2010,Nagy2011,Grujic2012,Lee2012,Kessler2012,Nissen2012,Kulaitis2013,Sieberer2013,Sieberer2014,Jin2014,LeBoite2014,Klinder2015,Naether2015}, topological phases \cite{Kapit2014,Zeuner2015,Peano2015} and quantum chaos \cite{Fernandez-Hurtado2014,Gao2015}. Open quantum lattices, especially with engineered coupling to baths, also play an increasingly vital role in the development of quantum information technology such as  quantum computing hardware \cite{Devoret2013a,Boixo2014,Schuld2014,Aron2014} and quantum networks \cite{Nickerson2014}.

The study of open quantum lattices tends to be challenging. Analytical and numerical techniques for open lattices are currently less developed than their counterparts for closed lattices. While for a large class of open quantum lattices the Lindblad master equation provides an appropriate theoretical framework \cite{Breuer2007,Alicki2007}, numerical methods for solving this master equation directly, such as exact diagonalization \cite{Briegel1993,Torres2014}, time evolution, or averaging of quantum trajectories \cite{Plenio1998}, are computationally demanding and become quickly infeasible as  lattice size increases. More sophisticated numerical techniques have been suggested and are further being developed, including matrix-product  \cite{Verstraete2004,Zwolak2004} and variational methods \cite{Weimer2015,Cui2015}. These methods can handle larger lattices to some degree, but come with their own specific drawbacks. 

As a result, the development of quantum simulators based on photons is particularly intriguing. Photonic systems represent an interesting open-system complement to the well-established paradigm of ultracold-atom quantum simulators. Since photons do not possess a chemical potential (however, see Ref.~\cite{Hafezi2014a} for a proposal to engineer a chemical potential), realistic photonic lattices will typically include coherent driving and photon loss \cite{Underwood2012}. Such systems will thus be a particular useful tool to better understand, gain intuition, and ultimately devise tractable effective models for open quantum lattices of interest.  Experiments with photonic quantum simulators will shed definitive light on both dynamical and steady-state phenomena by employing well-defined artificial lattice structures and systematically controlling parameters including drive strength, photon frequency, and strength of the mediated photon-photon interaction. Very promising experimental progress in this direction has already been made in the circuit QED architecture \cite{Houck2012a,Underwood2012,Schmidt2013}.

An experimental quantum simulator requires careful initial steps of \emph{validation}  \cite{Cirac2012} to ensure that the given physical system is correctly implementing the intended model. The validation procedure naturally demands that, for specific parameter regimes, theoretical understanding and reliable quantitative predictions are available and enable a comparison between theory and the experimental data obtained from the quantum simulator. 
\begin{figure*}
	\centering
	\includegraphics[width=\textwidth]{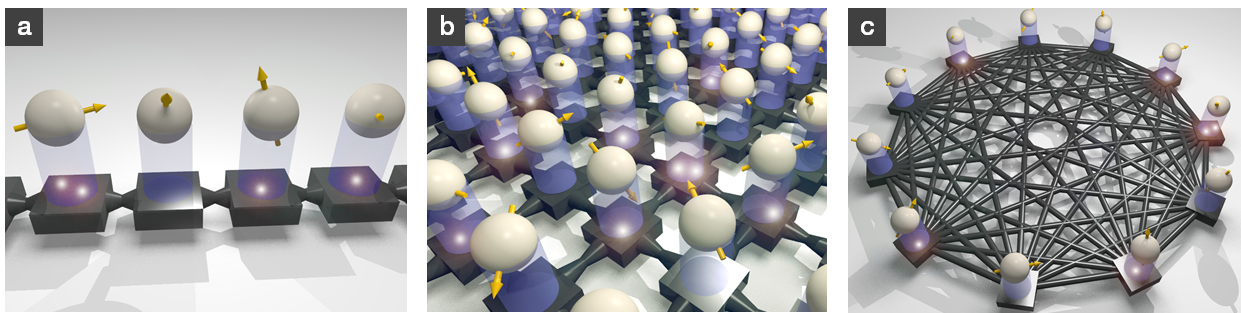}
	\caption{Open quantum lattices of different dimensionalities and geometries. The examples show Jaynes-Cummings lattices in which photons can hop between neighboring resonators (dark boxes) and experience an interaction mediated by the coupling to two-level systems (represented as spin systems). Lattice types of interest include (a) one-dimensional Jaynes-Cummings chains, (b) two-dimensional arrays such as the depicted square lattice, and (c) more artificial arrangements of theoretical interest, such as the global-coupling scenario where each site is connected to all other sites.
	}
	\label{fig:geometries}
\end{figure*}
For the purpose of validation, we utilize the well-controlled approximation scheme of Lindblad perturbation theory \cite{Garcia-Ripoll2009,Benatti2011, Kessler2012,Maizelis2014,Li2014}. We are particularly interested in the steady-state behavior of open lattices which can be directly related to experimental observables such as microwave transmission in circuit-QED lattices \cite{Houck2012a,Schmidt2013}. Steady-state quantities are of paramount importance for the detection of dissipative phase transitions \cite{Hur2008,Carusotto2009,Hartmann2010,Baumann2010,Nagy2011,Grujic2012,Kessler2012,Nissen2012,Kulaitis2013,Joshi2013,Sieberer2014,Ruiz-Rivas2014,LeBoite2014,Naether2015,Carmichael2015}.
In the work presented here, we take a crucial step beyond finite-order perturbation theory by demonstrating a partial resummation of the perturbation series. We then employ this method to study an open Jaynes-Cummings lattice [\cref{fig:geometries}] and establish that the resummation affords a significant improvement of the approximation accuracy.  We illustrate the method's versatility in handling both finite-size and infinite lattices as well as different geometries and dimensionalities in a natural way. The method is hence well suited for validating data from the first circuit-QED quantum simulators currently being investigated \cite{Houck_inprep}.

Our discussion is structured as follows. We set the stage  with a general review of (Markovian) open quantum lattices  in \cref{sec:background}, examining their theoretical description in terms of the Lindblad master equation (\ref{sec:open_quantum_lattices}), and the use of third-quantization methods \cite{Prosen2008,Prosen2010,Dzhioev2011} to obtain exact solutions for non-interacting open quantum lattices (\ref{sec:third_quantization}). \cref{sec:formalism} forms the centerpiece of our theoretical framework: here, we introduce the resummation scheme by which we can include certain perturbative corrections up to infinite order. 
We then apply this technique to the open Jaynes-Cummings (JC) lattice model in \cref{sec:JC_lattices}. After a few preparatory steps (\ref{sec:preparatory steps}), we perform the resummation and obtain results for steady-state observables (\ref{sec:perturbative_treatment}). We finally compare our results to both finite-order perturbation theory and exact solutions where available, and discuss the role of finite-size effects and lattice structures (\ref{sec:Result}). We conclude with a summary in Section \cref{sec:conclusion}.

\section{Background}
\label{sec:background}
\subsection{Lindblad formalism for driven, dissipative lattices}
\label{sec:open_quantum_lattices}
Open quantum lattice models are widely used to study many-body physics under nonequilibrium conditions. There exists a large variety of such lattice models, including open photon lattices \cite{Ozawa2014,Hafezi2014,Ningyuan2015}, Jaynes-Cummings lattices \cite{Knap2011,Nunnenkamp2011,Nissen2012,Grujic2012,Kulaitis2013,Ruiz-Rivas2014},  Dicke-type models \cite{Baumann2010,Nagy2011,Bhaseen2012,Konya2012,Zou2014} and lattices with different types of interactions \cite{Jin2013,Jin2014,Schiro2015}. Open lattices are not limited to bosons but may also involve fermions and spins, and may include both on-site and off-site interactions. We shall denote the Hamiltonians governing the unitary evolution from on-site and off-site terms by  $\mathsf{h}_{\vp}$ and $\mathsf{V}_{\vp \vp'}$, respectively. The resulting generic system Hamiltonian is then given by
\begin{equation}
\label{eq:hamform}
\sfH = \sum_{\vp} \mathsf{h}_{\vp} + \sum_{\vp\not= \vp'} \mathsf{V}_{\vp \vp'}.
\end{equation}
In many cases, off-site interactions can be limited to nearest-neighbor pairs $\lf< \vp , \vp' \rt>$.

For an open quantum lattice, we further account for the coupling to environmental degrees of freedom. Considering the effective dynamics of the lattice only, one finds that the environment generally induces non-unitary evolution which cannot be captured by an effective lattice Hamiltonian. The environment induces effects such as dissipation and decoherence, so that the time evolution of the reduced density matrix of the lattice $\rho$ generally deviates from the unitary von-Neumann equation $\dot{\rho} =  - i [\sfH,\rho]$.

In many relevant cases of weak coupling to the environment, the lattice system will undergo Markovian dynamics: the state of the system at the current time $t$ fully determines the  state at the slightly advanced time $t+dt$ in the future. (Another way to express this is to say that there are no ``memory effects.'')   Under a fairly general set of conditions \cite{Breuer2007}, the time evolution of $\rho$ is then described by the Lindblad master equation \cite{Breuer2007,Alicki2007},
\be
\label{eq:sL_general}
\dot{\rho}  = - i \lf[\sfH , \rho \rt] + \sum_{j}  \ga_j \sD \lf[ \mathsf{f}_j \rt]\rho.
\ee
The influence of the environment is thus encoded in the damping terms $\sum_{j}  \ga_j \sD [ \mathsf{f}_j ]$, where $\sD[\mathsf{f}_j]\rho \equiv \mathsf{f}_j \rho \mathsf{f}_j\dg - \mathsf{f}_j\dg \mathsf{f}_j \rho /2 - \rho \,\mathsf{f}_j\dg f_j /2$ is called the dissipator. Typically, each particular  jump operator $\mathsf{f}_j$ points to a particular non-unitary process. For example, photon decay commonly results from the photon annihilation operator $\mathsf{a}$ acting as the jump operator. (This statement is over-simplifying matters somewhat. In general, care must be taken to derive the appropriate jump operator for a particular system and environment coupling \cite{Cresser1992,Breuer2007}.) The prefactor $\ga_j$ characterizes the rate of the damping process. 

\begin{figure}
	\centering
	\includegraphics[width=0.7\columnwidth]{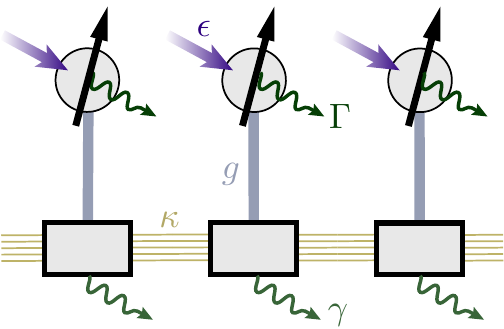}
	\caption{Constituents of the Jaynes-Cummings lattice: two-level systems are driven coherently with strength $\epsilon$, and exchange excitations with local harmonic oscillators at a rate set by $g$. Together, a pair of two-level system and oscillator form one site of the Jaynes-Cummings system. Nearest-neighbor sites are coupled by propagation of oscillator excitations with rate $\sim\kappa$. Dissipation is included in the form of two-level relaxation (rate $\Gamma$) and oscillator relaxation (rate $\gamma$). In the circuit-QED realization, two-level systems are implemented by superconducting qubits, and oscillators by photon-modes of on-chip microwave resonators with typical frequencies in the range of a few GHz.
	}
	\label{fig:lattice_structure}
\end{figure}

As a concrete example of an open quantum lattice and paradigm for interacting photon lattices, we consider the driven, damped Jaynes-Cummings lattice [\cref{fig:lattice_structure}]. In this system, each site consists of a harmonic oscillator, such as the mode of an electromagnetic resonator, coupled coherently to a two-level system, referred to as ``qubit'' in the following. 
Each site can be driven by a coherent tone. For simplicity, we consider the situation of a global drive frequency $\omega_d$, identical on each site.

The single-site Hamiltonian for this lattice is the Jaynes-Cummings Hamiltonian plus drive term,
\begin{align}
\label{eq:JCH}
\mathsf{h}_{\vp} &=  \dwrm{\vp} \mathsf{a}_{\vp}\dg \mathsf{a}_{\vp} + \dwq_{\vp} \spvp \smvp  +g_{\vp}  \lf(\mathsf{a}_{\vp} \spvp + \mathsf{a}_{\vp}\dg \smvp \rt)\\\nonumber
&\quad  + \epsilon_\vp \lf(\mathsf{a}_{\vp} + \mathsf{a}_{\vp}\dg  \rt).
\end{align}
In the usual way, we have already eliminated the original time dependence of the drive by switching to a frame rotating with the drive frequency. Consequently, the photon and qubit terms involve frequency detunings relative to the drive. Specifically, we have 
$\dwrm{\vp} \equiv \omega_{\vp} - \omega_d$ for the photon mode with frequency $\omega_{\vp}$, and $\dwq_{\vp} \equiv \Omega_{\vp} - \omega_d$ for the qubit with frequency $\Omega_{\vp}$ on site $\vp$. Photon and qubit excitations are created (annihilated) by the standard ladder operators $\mathsf{a}_{\vp}\dg$  and $\spvp$ ($\mathsf{a}_{\vp}$  and $\smvp$), respectively.
We denote the strengths of the Jaynes-Cummings coupling  and of the coherent drive tone by $g_{\vp}$ and $\epsilon_\vp$. Off-site terms of this lattice arise from the hopping of photons between sites $\vp$ and $\vp'$ with rate $\kappa_{\vp\vp'}$:
\be
\label{eq:hoppingH}
\mathsf{V}_{\vp\vp'} = \kappa_{\vp\vp'}(\mathsf{a}_{\vp}\dg \mathsf{a}_{\vp'} + \mathsf{a}_{\vp} \mathsf{a}_{\vp'}\dg). 
\ee
The full Hamiltonian consists of the appropriate sum of on-site and off-site contributions (as given above).

Finally, we consider the damping induced by photon decay (rate $\gar_\vp$) and qubit relaxation (rate $\gaq_\vp$). If both processes occur due to coupling to separate zero-temperature baths, then the appropriate jump operators can be shown to be the photon annihilation operator $\mathsf{a}_\vp$ and the pseudo-spin lowering operator $\smvp$. Overall, we thus obtain the Lindblad master equation
\be
\label{eq:rhoJC}
\dot\rho= -i \lf[\sfH ,\rho \rt] + \sum_{\vp} \gar_{\vp} \sD \lf[ \mathsf{a}_{\vp} \rt] \rho+ \sum_{\vp} \gaq_{\vp} \sD \lf[ \smvp \rt] \rho.
\ee

We will frequently find it convenient to write the Lindblad master equation in the short form 
$\dot{\rho} = \sL \rho$. 
Here, $\sL$ is the so-called Liouville superoperator. The term ``superoperators'' designates an object mapping an ordinary Hilbert-space operators such as $\rho$ to another Hilbert-space operator. (In our notation, we distinguish superoperators, operators, and real/complex numbers by using double-stroke, sans-serif, and regular lettering, respectively.) Using this shorthand of the master equation, we can easily characterize the steady state $\rs$ of the lattice: it is the stationary solution of this Lindblad master equation, i.e.\
\be
\dot{\rho}_\text{s} = \sL \rs = 0
\ee
with normalization $\Tr\rs=1$.

Mathematically, the equation $ \sL \rs = 0$ is a system of linear equations for the components of $\rs$. It can also be interpreted as a special instance of the eigenvalue problem for the Liouville superoperator $\sL$,
\be
\label{eq:sL_eigensystem_right}
\sL \evr{\mu} = \eva{\mu} \evr{\mu},
\ee
namely as the instance of eigenvalue $\lambda_\mu=0$. (We will always take $\mu=0$ to denote this case in the following.)
The steady state $\rs$ is thus the eigenstate $\evr{0} = \rs$ of $\sL$ associated with the eigenvalue $\eva{0} = 0$. While it is tempting to think of \cref{eq:sL_eigensystem_right} as a superoperator-analogue of the stationary Schr\"odinger equation, it is important to note that the Liouville superoperator $\sL$ is in general not hermitian, i.e.\ $\sL\dg$ is not equal to $\sL$. Hence, its eigenvalues may be complex-valued and
we have to distinguish right eigenstates $\evr{\mu}$ from left eigenstates $\evl{\mu}$,  given by
\be
\label{eq:sL_eigensystem_left}
\evl{\mu}\dg \sL = \evl{\mu}\dg \eva{\mu}.
\ee
As long as we keep this in mind, however, it is useful to mimic bra-ket notation and allow ourselves the  freedom to write operators and their adjoints also in the alternative form $\evr{\mu}\leftrightarrow |\evr{\mu})$ and $\evl{\mu}\dg\leftrightarrow (\evl{\mu}|$. We can then denote the Hilbert-Schmidt inner product between operators as $\sinner{\mathsf{x}}{\mathsf{y}}\equiv \sTrb{\mathsf{x}\dg \mathsf{y}}$. Linear algebra dictates that  
the right and left eigenstates of $\sL$ are orthogonal and, by appropriate normalization, can be chosen orthonormal,
\be
\label{eq:bi_orthonormal_condition}
\inner{\evl{\nu}}{\evr{\mu}} = \delta_{\mu \nu}.
\ee

Assuming that the eigenstates of $\sL$ form a complete set \footnote{The Liouville superoperator $\sL$ is not hermitian, so that completeness of the set of eigenstates is not guaranteed, though heuristically very common for $\sL$.}, we can represent an arbitrary  operator $\mathsf{X}$ as
\be
|\mathsf{X})=\sum_{\mu}|\evr{\mu})(\evl{\mu}|\mathsf{X})
=\sum_{\mu}(\mathsf{X})_\mu\, |\evr{\mu}),
\ee
and an arbitrary superoperator $\sA$ as 
\be
\sA = \sum_{\mu,\nu} |\evr{\mu})(\evl{\mu}|\sA |\evr{\nu})(\evl{\nu}| = 
\sum_{\mu,\nu} (\sA)_{\mu\nu}\,|\evr{\mu})(\evl{\nu}|.
\ee
Except for the matter of left vs.\ right eigenvectors, these expressions are familiar from the usual decomposition of states and operators in Hilbert space, and we will make use of them in \cref{sec:formalism}.

\subsection{Non-interacting open lattices, third-quantization method}
\label{sec:third_quantization}
Due to competition of on-site interaction and off-site hopping, as well as non-conservation of excitation number, the open Jaynes-Cummings lattice is not exactly solvable in general. The model does become tractable once the coupling to qubits is removed, by which one obtains an open lattice of non-interacting photons subject to coherent driving and photon loss (and a disconnected set of qubits). Third-quantization methods can then be used to find the exact solutions to the stationary Lindblad equation \eqref{eq:sL_eigensystem_right} \cite{Prosen2008,Prosen2010,Dzhioev2011}. To sketch the procedure, we consider a uniform photon-hopping Hamiltonian without disorder, 
$
\sfH_\text{P} =  \sum_{\vp} \lf[ \dwrm{} \mathsf{a}_{\vp}\dg \mathsf{a}_{\vp} + \ep (\mathsf{a}_{\vp}\dg + \mathsf{a}_{\vp} ) \rt]
+ \ka\sum_{\lf<\vp, \vp' \rt>}  (\mathsf{a}_{\vp}\dg \mathsf{a}_{\vp'} + \mathrm{h.c.})
$
and the Liouville superoperator
$
\sL_\text{P} {}_\bullet= -i \lf[ \sfH_\text{P}, {}_\bullet \rt]+ \gar\sum_{\vp}  \sD \lf[ \mathsf{a}_{\vp} \rt] {}_\bullet
$.
We bring the hopping Hamiltonian into diagonal form in two steps. First, we introduce the appropriate Bloch states which are labeled by quasimomentum $\vk$, and describe the normal modes of the undriven photon lattice $\sfH_\text{P}|_{\ep=0}$. Second, we note that uniform driving is synonymous with driving of the $\vk=0$ mode. Hence, we can eliminate the drive term by a coherent displacement of this mode, $\mathsf{a}_{\vk=0}\to \mathsf{a}_{\vk=0}+\alpha$, where $\alpha$ depends on both drive strength and relaxation rate. These steps yield the simplified Hamiltonian $\sfH_\text{P}' =  \sum_{\vk} \dwrm{\vk} \mathsf{a}_{\vk}\dg \mathsf{a}_{\vk}$ and Liouville superoperator 
$
\sL_\text{P}' {}_\bullet= -i \lf[ \sfH_\text{P}, {}_\bullet \rt]+ \gar\sum_{\vk}  \sD \lf[ \mathsf{a}_{\vk} \rt] {}_\bullet
$.

This Liouville superoperator can now be diagonalized analytically by using the third-quantization method \cite{Prosen2008,Prosen2010,Dzhioev2011} (or alternative techniques, e.g., \cite{Buchleitner2002}). We employ the superoperators $\sba{\vk}$, $\sbap{\vk}$, $\sbb{\vk}$, $\sbbp{\vk}$, which mimic  boson annihilation and creation operators and are defined by:
\begin{align}
\label{eq:def_sb}
\sba{\vk} \rho &= \mathsf{a}_{\vk} \rho \quad\text{\ \ and\ \ }\quad \sbap{\vk} \rho = \mathsf{a}_{\vk}\dg \rho - \rho \mathsf{a}_{\vk}\dg, \\
\sbb{\vk} \rho &= \rho \mathsf{a}_{\vk}\dg  \quad\text{\ \ and\ \ }\quad \sbbp{\vk} \rho = \rho \mathsf{a}_{\vk} - \mathsf{a}_{\vk} \rho.
\end{align}
While $\sbap{\vp}$ and  $\sba{\vp}$ are not proper adjoints, the use of the unconventional ``$\dda$'' symbol is motivated by the the fact that it leads to commutation relations of the ordinary form,
\begin{align}
\label{eq:almost_CCR}
&[\sba{\vk}, \sbap{\vk'} ] = [\sbb{\vk}, \sbbp{\vk'} ]
 =   \delta_{\vk \vk'}, 
\end{align}
and all other commutators vanish.
Thanks to this commutator algebra, $\sL_\text{P}'$ takes on the compact form \cite{Prosen2010},
\be
\label{eq:general_diagonalized_form}
\sL_\text{P}'  =  \sum_{\vk} \lf( \sNmaev{\vk} \sbap{\vk} \sba{\vk} +\sNmaev{\vk}^* \sbbp{\vk} \sbb{\vk} \rt).
\ee
Here, $\sba{\vk}$ ($\sbap{\vk}$) and $\sbb{\vk}$ ($\sbbp{\vk}$) may be viewed as ``normal-mode'' superoperators, and complex-valued ``mode frequencies'' 
$\sNmaev{\vk} = -i \, \dwrk -\gar/2$ and $\sba{\vk}^*$.

From this, it is straightforward to read off eigenvalues and eigenstates of $\sL_\text{P}'$, analogous to the way we find eigenvalues and eigenstates of a non-interacting boson  Hamiltonian. For a given $\vk$-mode, the right and left ``vacuum states'' obey  $\sba{\vk} \evrs{\vk} = \sbb{\vk} \evrs{\vk} = 0$ and $\evlr{\vk}{0}{0}\sbap{\vk} = \evlr{\vk}{0}{0}\sbbp{\vk}  = 0$. The right ``vacuum state'' is therefore the projector $\evrs{\vk}=|0_\vk\rangle \langle 0_\vk|$ onto the pure state without any photons in mode $\vk$. One can show that the left ``vacuum states'' must always coincide with the identity operator, $\evlr{\vk}{0}{0} = \mathsf{1}$. Once the coherent displacement is reversed, we thus obtain  $\rho_\text{s}=\bigotimes_{\vk\not=0}|0_\vk\rangle \langle 0_\vk|\otimes |\alpha_{\vk=0}\rangle \langle \alpha_{\vk=0}|$ for the steady state of the original Liouville superoperator $\sL_\text{P}$.

The ``excited'' eigenstates of $\sL_\text{P}'$ are obtained by acting with the creation superoperators on the ``vacuum states.'' For given $\vk$, this means
\be
\label{eq:sL_P_right_eigen}
\evrr{\vk}{m}{n} = \frac{1 }{\sqrt{m! n!}} (\sbap{\vk})^m (\sbbp{\vk})^n \evrs{\vk}\,,
\ee
and
\be
\label{eq:sL_P_left_eigen}
\evlr{\vk\dag}{m}{n} = \frac{ 1 }{\sqrt{m! n!}} \evlr{\vk\dag}{0}{0}(\sba{\vk})^m (\sbb{\vk})^n\,.
\ee
The corresponding eigenvalues are $\evar{\vk}{m}{n} = m \sNmaev{\vk}+ n  \sNmaev{\vk}^* $. When forming the appropriate product states and summing eigenvalues over $\vk$, we thus obtain the entire spectrum of the Liouville superoperator.

While our brief review here has been limited to the simplest scenario, we emphasize that the third-quantization method applies to a much larger class of lattice models which may involve bosons, fermions as well as spins  \cite{Prosen2008a,Prosen2011,Pizorn2013,Medvedyeva2015}.

\section{Lindblad Perturbation Theory and Resummation}
\label{sec:formalism}
In this section, we present the formalism of Lindblad perturbation theory and its resummation. This section remains general and is applicable to Markovian open quantum systems of various types. The concrete application of the formalism to an open Jaynes-Cummings lattice follows in \cref{sec:JC_lattices}. Together, these two sections constitute the central result of our paper.


\begin{figure}
	\centering
	\includegraphics[width=0.9\columnwidth]{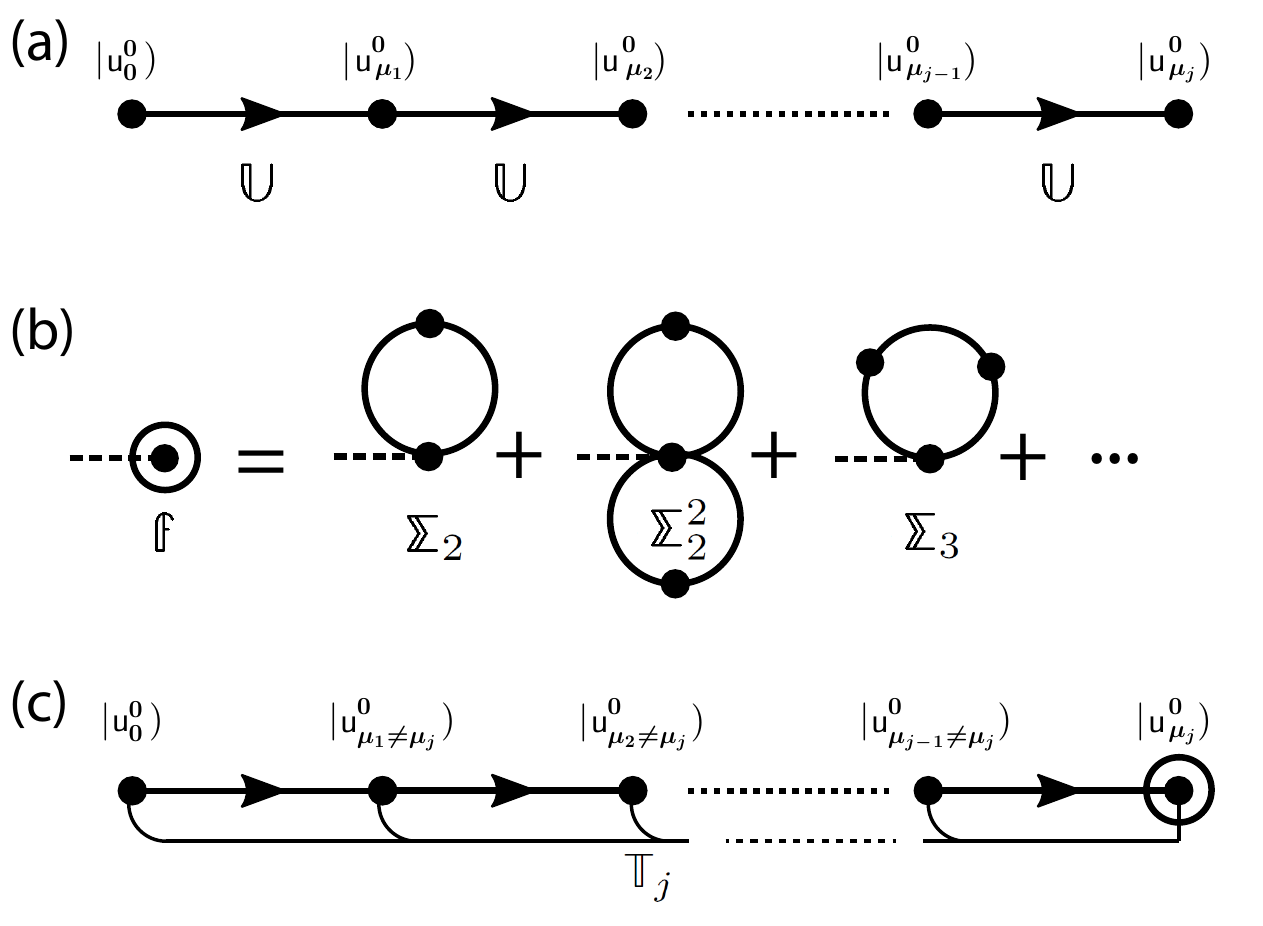}
	\caption{Diagrams for perturbative corrections.
\textbf{(a)} The order-$j$ correction to the  steady state [\cref{eq:rsc_recursive_2}] is depicted as a chain of $j$ lines, each representing one factor $\sU$. The chain connects $j+1$ dots symbolizing $\sL_0$-eigenstates. The leftmost state is the unperturbed steady state  $|\rsoc{j})=|\evrc{0}{0})$, the rightmost one the unperturbed eigenstate $|\evrc{\mu_j}{0})$,  giving rise to one specific correction term. The full correction is obtained by summation over final and intermediate states $\mu_1,\ldots,\mu_j$.
\textbf{(b)} The resummation superoperator $\spf{}$ is the sum of all (reducible and irreducible) $\sloop$-superoperators. Each  $\sloop$-diagram starts and ends with the same unperturbed eigenstate. 
\textbf{(c)} Resummation combines evaluation of $\Sigma$ and $\sUp$-superoperators. Terms of rank $j$ are comprised of the fully  off-diagonal chain specified by $\sUp_j$ and final application of the resummation superoperator.
	}
	\label{fig:general_PT_diagram}
\end{figure}

Consider the general case of an open quantum system with Hamiltonian $\sfH$ and Liouville superoperator $\sL$. We shall assume that $\sL$ is amenable to a perturbative treatment and can be decomposed into a sum $\sL=\sL_0 + \sL_1$, consisting of the unperturbed Liouville superoperator $\sL_0$, and the perturbation $\sL_1$. For $\sL_0$ to qualify as such, it is expected that we can obtain its spectrum exactly. We denote the resulting unperturbed eigenvalues by $\evac{\mu}{0}$, and the corresponding unperturbed right and left eigenstates by $|\evrc{\mu}{0})$ and $(\evlc{\mu}{0}|$, respectively.


The spectra of $\sL$ and $\sL_0$ differ, but corrections may be calculated by a perturbative series expansion if $\sL_1$ is ``sufficiently small.''  The corrections to eigenvalues and eigenstates can then be determined recursively, order by order \cite{Li2014}.
Our interest here primarily regards the steady state $\rs$, and we apply Lindblad perturbation theory assuming the non-degenerate case in which the steady state is unique. Two remarks may be useful for clarification. First, we emphasize that non-degeneracy refers to the spectrum of $\sL_0$, not to the Hamiltonian $\sfH$; we make no assumptions about the spectrum of $\sfH$. Second, we note that non-uniqueness of the steady state and resulting non-analyticities are crucial at the phase boundary of a dissipative phase transition. Perturbative series expansions will generally not hold directly at such a boundary, but may still be applicable in its vicinity.

Turning now to the concrete series expansion $|\rs)=\sum_j |\rsoc{j})$ of the steady state, we note that the $j$-th order contribution $\rsoc{j}$  is obtained from the recursion relation
\be
\label{eq:rsc_recursive}
|\rsoc{j}) =- \sL_0\ginv \sL_1 |\rsoc{j-1}).
\ee
Here, inversion of $\sL_0$ will always be understood as restricted to the space orthogonal  \footnote{If the unperturbed eigenstates are not complete, a generalized inverse \cite{Benatti2011,Li2014} or projection techniques \cite{Garcia-Ripoll2009,Kessler2012} must be used.} to the unperturbed steady state, i.e.,
$
\sL_0\ginv = \sum_{\mu>0} ( \evac{\mu}{0} )^{-1} |\evr{\mu})(\evl{\mu}|.
$
With this, we obtain the formal expression
\be
 |\rsoc{j})=\lf(- \sL_0\ginv \sL_1\rt)^j |\rsoc{0}) =\sU^j |\rsoc{0})  ,
\ee
where $\rsoc{0}$ is the unperturbed steady state of $\sL_0$, and we have introduced the shorthand $\sU \equiv - \sL_0\ginv \sL_1$. The matrix elements of the $\sU$-superoperator are
\be
(  \sU )_{\mu \nu}  = \sinner{ \evlc{\mu}{0} }{  \sU| \evrc{\nu}{0} } = -  \sinner{ \evlc{\mu}{0} }{  \sL_1 |\evrc{\nu}{0} }/\evac{\mu}{0}
\ee
for $\mu>0$ and $(  \sU )_{0 \nu}=0$. Using this shorthand, we write the $j$-th order contribution to the steady state in the form of the chain expression
\begin{equation}
\label{eq:rsc_recursive_2}
|\rsoc{j}) = \sum_{ \mu_1, \mu_2 , \cdots, \mu_j}  |\evrc{\mu_j}{0}) \cdot \lf( \sU \rt)_{\mu_j \mu_{j-1}} \cdots\lf( \sU \rt)_{\mu_{2} \mu_{1}} \lf( \sU \rt)_{\mu_{1} 0},
\end{equation}
and  represent it diagrammatically as shown in  \cref{fig:general_PT_diagram}(a). As a diagrammatic rule, we choose dots to represent unperturbed eigenstates, $|\evrc{\mu_j}{0})$, and interconnecting lines to represent factors of $\sU$. Reading from the left to right, the leftmost state is the unperturbed steady state $|\rsoc{0}) = | \evrc{0}{0})$, and the rightmost one the final state $|\rsoc{\mu_j})$ which appears explicitly in the expression \eqref{eq:rsc_recursive_2}. All intermediate states and the final state, $\mu_1$ through $\mu_j$  are subject to summation, but cannot coincide with the initial unperturbed steady state due to  $(  \sU )_{0 \nu}=0$.


To facilitate our resummation scheme for the steady-state series
\be
\label{eq:rs_sum_1}
|\rs)  =\sum_{j=0}^{\infty} \sU^j  |\rsoc{0}),
\ee
we decompose the $\sU$-superoperator into two parts
\be
\label{eq:U1_decomp}
\sU^1 = \sloopn{1} + \sUpc{1}.
\ee
To make the definition parallel to expressions to come, we have explicitly recorded the exponent $j=1$ on the left-hand side. We now specify the terms on the right-hand side in such a way that resummation takes a particular simple form.
We define the first-order $\sloopn{}$-superoperator as the diagonal part of $\sU^1$, i.e.\ $( \sloopn{1} )_{\mu \nu} =\delta_{\mu \nu} ( \sU^1 )_{\mu \nu}$. Accordingly, $\sloopn{1}$ and $\sL_0$ share the same set of right eigenstates, i.e., $\sloopn{1} |\evrc{\mu}{0}) = \evasloop_{1;\mu} |\evrc{\mu}{0})$, and the eigenvalue is $\evasloop_{1;\mu} = ( \sU^1 )_{\mu \mu}$. We will see that this simplicity of $\sloopn{1}$ will be important for the resummation of the series,  \cref{eq:rs_sum_1}. The term $\sUpc{1}$ in \cref{eq:U1_decomp} is the off-diagonal remainder of the $\sU$-superoperator.

A simplification of the previous expressions occurs when making the natural assumption that the perturbation $\sL_1$ is itself off-diagonal with respect to the unperturbed eigenstates of $\sL_0$. (Whenever $\sL_1$ does not satisfy this assumption, a simple re-definition of $\sL_0$ and $\sL_1$ can be used to turn $\sL_1$ off-diagonal.) Now, if $\sL_1$ is off-diagonal, so is $\sU=-\sL_0^{-1}\sL_1$ and we immediately obtain
\be
\label{eq:def_SE_0}
\sloopn{1} = 0 \text{ \ and \ } \sUpc{1} = \sU.
\ee
This may initially make the decomposition of $\sU$ seem pointless, but we will see momentarily that this simplification does not carry over to higher orders $j>1$, thus justifying our approach.

We next consider the second-order term $\sU^2|\rsoc{0})$, which warrants a decomposition of $\sU^2=\sUpc{1} \sU$ into
\be
\label{eq:sU2_separation}
\sUpc{1} \sU = \sloopn{2}+ \sUpc{2}.
\ee
Analogous to our strategy above, we define the second-order $\sloopn{}$-superoperator, as the diagonal part of the left-hand side,
\be
\label{eq:L2_definition}
\lf(\sloopn{2}\rt)_{\mu \nu} =  \delta_{\mu \nu} \lf(\sUpc{1} \sU \rt)_{\mu \nu}
=\delta_{\mu \nu}\sum_{\tau}  
 \lf(\sU \rt)_{\nu \tau} \lf(\sU \rt)_{\tau \nu}.
\ee
Note that the off-diagonal character of $\sU$ automatically leads to exclusion of the term with $\tau=\nu$.
As before, $\sUpc{2}$ represents the remaining off-diagonal part in \cref{eq:sU2_separation}.
We represent $\sloopn{2}$ by the loop diagram shown in \cref{fig:general_PT_diagram}(b). Since $\sloopn{2}$ is diagonal, its initial and final state $|\evrc{\mu}{0})$ must be identical. However, the intermediate state $|\evrc{\tau}{0})$ involved in the expression \eqref{eq:L2_definition} must differ from $|\evrc{\mu}{0})$.

For resummation of terms to infinite order, we need to formulate our decomposition strategy for arbitrary order $j$. It is natural to extend the definitions for diagonal and off-diagonal superoperators, \cref{eq:sU2_separation,eq:L2_definition}, by setting
\begin{align}
\label{eq:separation_SE}
\sUpc{j-1} \sU  = \sloopn{j}+ \sUpc{j}, \qquad
\lf(\sloopn{j}\rt)_{\mu \nu} = \delta_{\mu \nu} \lf( \sUpc{j-1} \sU \rt)_{\mu \nu}.
\end{align}
The recurrence relation is solved by 
\be
\label{eq:TT}
\sUpc{j} = \llbracket\llbracket\cdots\llbracket\llbracket \sU\rrbracket \sU\rrbracket\cdots \sU\rrbracket\sU\rrbracket \qquad \text{($j$ times)},
\ee
where $\llbracket \mathbb{A} \rrbracket$ denotes the off-diagonal part of $\mathbb{A}$ with respect to the unperturbed basis $\{|\evrc{\mu}{0})\}$. 
We must note, however, that the definition \eqref{eq:separation_SE} does not yet  determine a unique separation scheme beyond second order. Consider, for instance, the case of the third-order term involving $\sU^3$. While we know the decomposition of $\sU^2=\sUpc{1}\sU$ from \cref{eq:sU2_separation}, we still have the freedom to perform the substitution for either $\sU^3=\sU(\sU^2)$ or $\sU^3=(\sU^2)\sU$. Both forms are mathematically equivalent, but only the systematic usage of one replacement rule produces expressions for which resummation becomes simple. We will consistently employ the form 
\be
\label{eq:sU_decomposition_rule}
\sU^j = \underbracket[0.75pt][2pt]{\sU^{j-1}} \sU,
\ee
where $\underbracket[0.75pt][2pt]{\sU^{j-1}}$ signals that $\sU^{j-1}$ is to be replaced by an expression composed of  $\sloopn{}$,  $\sUpc{}$, and $\sU$-superoperators. Multiple replacements, in some cases making use of the identity $\sU=\sUpc{1}$, may be necessary to reach the final decomposed form  only involving $\sloopn{}$ and $\sUpc{}$-superoperators.

For illustration, we consider the decompositions of $\sU^3$, $\sU^4$, and $\sU^5$. For the third-order case, we first make use of \cref{eq:sU_decomposition_rule} and then \cref{eq:separation_SE} to obtain
\be
\sU^3 = \underbracket[0.75pt][2pt]{\sU^2} \sU =  \underbracket[0.75pt][2pt]{\sUpc{1}\sU} \sU=\sUpc{2} \sU + \sloopn{2} \sU.
\ee
The last term on the right-hand side cannot be simplified further (except for substituting $\sU=\sUpc{1}$), the first term is further decomposed by using \cref{eq:separation_SE}, leading to the final expression
\be
\label{eq:sU3_separation}
\sU^3 = \underbracket[0.75pt][2pt]{\sUpc{2} \sU} + \sloopn{2} \sUpc{1} = \sloopn{3} + \sUpc{3} +\sloopn{2} \sUpc{1}.
\ee
For the fourth order, we merely sketch the decomposition,
\begin{align}
\nonumber
\sU^4  &= \underbracket[0.75pt][2pt]{\sU^3}\sU =  
\sloopn{2}\underbracket[0.75pt][2pt]{\sUpc{1}\sU} +  \sloopn{3}\sU + \sloopn{4} + \sUpc{4}\\
&=\sloopn{4} + \sloopn{2}^2 +\sloopn{3} \sUpc{1} + \sloopn{2} \sUpc{2}  +\sUpc{4}.
\label{eq:sU4_separation}
\end{align}
We give the fifth-order result without showing substeps,
\begin{align}
\label{eq:sU5_separation}
\sU^5  = & \  \sloopn{5} + \sloopn{2} \sloopn{3} + \sloopn{3} \sloopn{2}   \nn\\
& + \lf( \sloopn{4} + \sloopn{2}^2 \rt) \sUpc{1} + \sloopn{3} \sUpc{2} + \sloopn{2} \sUpc{3} + \sUpc{5}.
\end{align}

Inspection of \cref{eq:sU3_separation,eq:sU4_separation,eq:sU5_separation} indicates a systematic structure underlying the expressions, namely
\be
\label{eq:pattern}
\sU^j = \sum_{k=0}^{j} \spSE{j-k} \sUpc{k}.
\ee
Each term in this sum has one factor of  $\sUpc{k}$ of order $0\le k \le j$ and a prefactor $\spSE{j-k}$ consisting of all possible combinations of $\sloopn{}$-superoperators  of total order $j-k$. A formal proof of this is given in \cref{app:proof_SE}. (Recall from \cref{eq:def_SE_0} that $\sloopn{1}=0$, which reduces the number of terms significantly.) 
Using the decomposition \eqref{eq:pattern} and regrouping terms according to each occurrence of $\sUpc{j}$, we can now rewrite the perturbation series for the steady state in the form
\be
|\rs) = \sum_{j=0}^{\infty}\sum_{k=0}^{j} \spSE{j-k} \sUpc{k}|\rsoc{0})
=\spf{}\, \sum_{j=0}^{\infty} \sUpc{j} |\rsoc{0}).
\ee
Here, the superoperator $\spf{}=\spf{}(\sloopn{1},\sloopn{2}, \cdots)=\sum_{n=0}^\infty\spSE{n}$ implements the resummation of terms that we have been aiming for. It is given by
\be
\spf{}= \openone + \sloopn{2} + \sloopn{3} +\sloopn{4} + \sloopn{2}^2 + 
\sloopn{5} + \sloopn{2} \sloopn{3} + \sloopn{3} \sloopn{2}
\cdots,
\ee
i.e., the sum of all possible products of $\sloopn{}$-superoperators (here explicitly shown up to fifth order). Diagrammatically, we represent $\spf{}$ in the form shown in  \cref{fig:general_PT_diagram}(b). Due to the definitions of $\sloopn{j}$ and $\sUpc{j}$ as diagonal and off-diagonal superoperators,  $\sloopn{j}$ corresponds to a loop diagram with  initial and final state being identical, all $(j-1)$ intermediate states being different from the initial/final state, and consecutive intermediate states being different from each other. 

The role of $\sloopn{j}$ resembles that of an irreducible self-energy contribution of order $j$ in closed-system perturbation theory. Indeed, if we define $\selfe = \sum_{j=1}^{\infty} \sloopn{j}$ as the sum of all irreducible ``self-energy'' contributions, we can rewrite 
$
\spf{} =  \sum_{j=0}^{\infty} \selfe^j = ( \openone -  \selfe )^{-1},
$
and obtain 
\be
\label{eq:resummed}
|\rs)  =\sum_{j=0}^{\infty} \frac{1}{ \openone -  \selfe }\sUpc{j}  |\rsoc{0})
\ee
for our resummed series expansion of the steady state. Due to the $(\openone-\selfe)^{-1}$ prefactor, each term $|\rsc{j}) = (\openone-\selfe)^{-1}\sUpc{j}  |\rsoc{0})$ in this sum includes perturbative corrections up to infinite order. We therefore call the term $|\rsc{j})$ the rank-$j$ term of the resummed series.  We note that, formally, \cref{eq:resummed} is an exact expression for the steady state. Practical calculations will typically involve a truncation in both the maximum summation index $j$ as well as the maximum order of irreducible self-energy contributions taken into account.

We represent individual terms $|\rsc{j})$ in the resummed series by the type of diagram shown in  \cref{fig:general_PT_diagram}(c). The final-state dot is marked with a circle to indicate the inclusion of the self-energy correction. The diagrammatic rules are similar to the case without resummation, \cref{fig:general_PT_diagram}(a), except that the off-diagonal nature of $\sUpc{j}$ in addition requires that all intermediate states be different from the final state. This is simple to infer when writing $\sUpc{j}$ in the form of \cref{eq:TT}.
 Each diagram then translates to an expression with the following structure:
\begin{align}
\label{eq:rsc_recursive_3}
|\rsc{j}) = &
\sum_{\mu_j}\sum_{\nu_1,\ldots,\nu_{j-1}\not=\mu_j}
|\evrc{\mu_j}{0}) \lf(\frac{1}{\openone - \selfe}\rt)_{\mu_j \mu_j}\\\nonumber
&\qquad\times  \lf( \sU \rt)_{\mu_j \nu_{j-1}} \lf( \sU \rt)_{\nu_{j-1} \nu_{j-2}} \cdots\;\lf( \sU \rt)_{\nu_{2} \nu_{1}} \lf( \sU \rt)_{\nu_{1} 0}.
\end{align}

\section{Application to the Open Jaynes-Cummings Lattice}
\label{sec:JC_lattices}
Lindblad perturbation theory and resummation as discussed in the previous section are applicable to a large class of open quantum systems. Here, we present its concrete use in studying the open Jaynes-Cummings lattice [\cref{eq:hamform,eq:JCH,eq:hoppingH,eq:rhoJC}] as a specific example of an open driven quantum lattice. The example is of particular interest due to its role as a minimal model for highly anticipated experiments with circuit-QED lattices \cite{Houck_inprep}. The photonic backbone of the lattice has already been demonstrated in the experimental work by Underwood et al. \cite{Underwood2012}.

\subsection{Preparatory steps}
\label{sec:preparatory steps}
We shall consider a uniform lattice, in which resonator frequencies, qubit frequencies, and related quantities have uniform values across the lattice. (Disorder levels, especially in qubit frequencies, may need to be considered carefully for detailed modeling of future circuit-QED experiments, but this consideration is beyond the scope of the present paper.) We find the modes of the photonic lattice structure by diagonalizing the $N\times N$ hopping matrix, $N$ being the number of lattice sites \footnote{For an infinite system, $N\gg1$ is the auxiliary lattice size involved in applying Born-von Karmann periodic boundary conditions.}. For periodic lattices, diagonalization is achieved in the usual way by switching from real space to momentum space via the transformation $\msar = \sum_{\vk} \at_{\vk}e^{i \vk \cdot \vp} /\sqrt{N}$. Here, photons inside the mode with quasi-momentum $\vk$ are annihilated by $\at_{\vk}$, and $\vk$ runs over all reciprocal lattice vectors from the first Brillouin zone. Note that $\vk = \vm$  corresponds to the uniform mode with identical amplitudes on all sites, which is the mode being excited by a global coherent drive. 

Depending on the values of model parameters, it is beneficial to perform a displacement transformation which eliminates the coherent drive on the photon mode and converts it into an effective qubit drive instead. This is particularly helpful when the uniform photon mode is approximately in a coherent state with a large number of photons. The coherent displacement then serves as a tool incorporating this insight directly into the unperturbed Liouville superoperator and mitigates the need for large photon number cutoffs. (In a regime of low photon occupation, however, the displacement transformation can be skipped and the perturbative treatment carried out directly.) The displacement is applied to the $\vk=\vm$ mode, i.e., 
$
\bt_{\vm} = \at_{\vm} - \alp ,
$
with 
$
\alp =-\sqrt{N}  \epsilon/(\dwrm{\vk=\vm} - i \gar/2)
$.

After this displacement, the resonator drive is converted into an effective qubit drive of strength $\epe = g \alp /\sqrt{N} $. The resulting Hamiltonian has the form
\begin{align}
\label{eq:displaced_H}
\sfH' & = \sum_{\vk} \sfH_{r}^{\vk} + \sum_{\vp} \sfH_{q}^{\vp} +\sum_{\vk,\vp} \sfH_{rq}^{\vk \vp},
\end{align}
where the three terms correspond to the photon, qubit, and photon-qubit coupling contributions.
The resonator part now lacks the drive term, $\sfH_{r}^{\vk} = \dwrk \bt_{\vk}\dg \bt_{\vk}$. (We define $\bt_{\vk} = \at_{\vk}$ for $\vk \neq \vm$ to unify notation.)
The qubit Hamiltonian including the effective drive reads 
\begin{equation}
\label{eq:displaced_qubit}
\sfH_{q}^{\vp} = \dwq\, \spvp \smvp + 
(\epe  \spvp + \epe^* \smvp ),
\end{equation}
and the
interaction Hamiltonian is given by
\begin{align}
\label{eq:H1_JC}
\sfH_{rq}^{\vk \vp} =
&  \frac{g}{\sqrt{N}}  \lf(  \bt_{\vk} \spvp  e^{i \vk \cdot \vp} +   \bt_{\vk}\dg \smvp e^{-i \vk \cdot \vp} \rt).
\end{align}
Finally, the dissipator term simply transforms as $\gamma\sum_{\vp} \sD \lf[ \at_{\vp} \rt]=\gamma\sum_{\vk} \sD \lf[\bt_{\vk} \rt]$.

In the absence of the interaction $\sfH_{rq}^{\vk \vp}$,  resonator modes and qubits decouple, and the associated master equation is exactly solvable. This presents us with an ideal starting point for a perturbative treatment of $\sfH_{rq}^{\vk \vp}$, which physically is a very sensible treatment of the dispersive regime.
The unperturbed Liouville superoperator is then $\sL_0 = \sum_{\vk} \sLr{\vk} +\sum_{\vp} \sLq{\vp}$ with separate photon contribution, $ \sLr{\vk} {}_\bullet = -i \sCom{\sfH_r^{\vk}} + \sD \lf[\bt_{\vk} \rt]{}_\bullet$, and qubit contribution, $ \sLq{\vp} {}_\bullet\equiv -i \sCom{\sfH_q^{\vp}} + \sD \lf[ \smvp\rt]{}_\bullet$. 

First, the photonic part $\sLr{\vk}$ is directly amenable to the diagonalization procedure we discussed in \cref{sec:open_quantum_lattices}. The resulting eigenvalues read $\evar{\vk}{m}{n} = m \sNmaev{\vk}+ n  \sNmaev{\vk}^*$ with $\sNmaev{\vk} = -i \, \dwrk -\gar/2$ ($m,n=0,1,\ldots$). The associated right and left eigenvectors $|\evrr{\vk}{m}{n})$ and $(\evlr{\vk}{m}{n}|$ are those given in \cref{eq:sL_P_right_eigen,eq:sL_P_left_eigen}.
Second, the qubit Liouvillian $\sum_{\vp} \sLq{\vp} $ can be diagonalized exactly since it decomposes into a direct product of $4\times4$ matrices. For each $\sLq{\vp}$, we denote the eigenvalues, and right/left eigenstates  by $\evaqvp{\mu}$, $|\evrqvp{\mu})$ and $(\evlqvp{\mu}|$, respectively ($\mu=0,\ldots,3$). Except for special parameters, analytical expressions for these quantities are too lengthy to provide much insight, and will hence not be recorded here. 

Altogether, right and left eigenstates of the full-lattice Liouvillian 
 $\sL_0$ thus take the form
\be
|\evrc{\vec{m}\vec{n}\vec{\mu}}{0}) = \bigotimes_{\vk} |\evrrvk{m_{\vk}}{n_{\vk}})  \bigotimes_{\vp} |\evrqvp{\mu_{\vp}}),
\ee
with corresponding eigenvalues
\be
\evac{\vec{m}\vec{n}\vec{\mu}}{0} = \sum_{\vk} \evarvk{m_{\vk}}{n_{\vk}}  + \sum_{\vp} \evaqvp{\mu_{\vp}}.
\ee
The multi-indices $\vec{m}$, $\vec{n}$, and $\vec{\mu}$ collect the sets of all ``quantum numbers'' $m_{\vk}$, $n_{\vk}$ and $\mu_{\vp}$.
With this, we are ready for the perturbative treatment of the Liouvillian $\sL_1$
capturing the Jaynes-Cummings interaction $\sfH_{rq}^{\vk \vp}$.

\begin{figure*}
	\centering
	\includegraphics[width=1\textwidth]{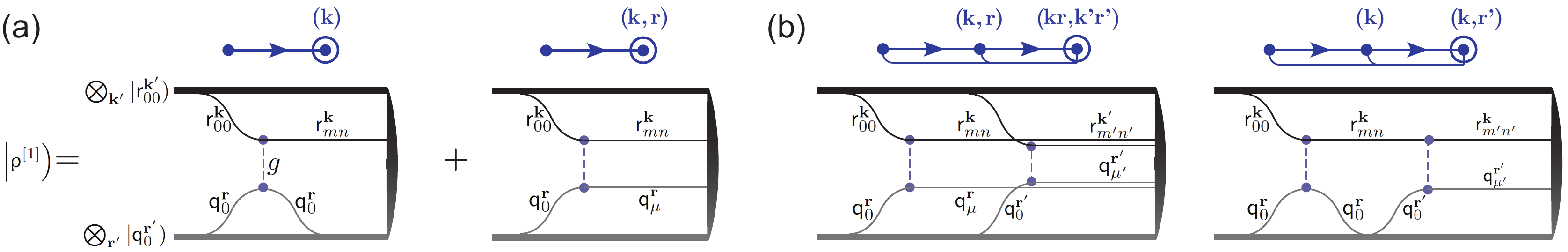}
	\caption{
		Diagrams for perturbative treatment and resummation of a  JC lattice. In each panel, the top part shows the general diagram labeled by constituents deviating from the steady-state configuration. Bottom parts are JC-lattice specific diagrams, with the upper branch denoting photon modes and the lower one qubit degrees of freedom. \textbf{(a)} Rank-1 corrections. There are two classes of terms with either a photon mode or a cluster of a photon mode $\vk$ and a qubit at $\vp$ deviating from the unperturbed steady-state configuration.Each interaction vertex $g$ must switch the photon mode configuration, but may leave that of the qubit unchanged. Terminating symbols on the right signal application of the resummation superoperator $(\openone-\sloop)^{-1}$. 
		\textbf{(b)} Two examples of rank-2 corrections, which  differ in the number of involved photon modes and qubits, and number of constituents deviating from the steady-state configuration. 
	}
	\label{fig:diagram_representation}
\end{figure*}
\subsection{Perturbative treatment and resummation}
\label{sec:perturbative_treatment}

The perturbation superoperator decomposes into  a sum $\sL_1 =\sum_{\vk,\vp}  \sL_1^{\vk \vp}$, in which each term  describes the interaction between an individual resonator mode $\vk$ and a qubit at position $\vp$: 
\begin{align}
\label{eq:JC_perturbation}
\sL_1^{\vk \vp}{}_\bullet
=  -i\frac{g}{\sqrt{N}}
\sCom{\bt_{\vk} \spvp  e^{i \vk \cdot \vp} +   \bt_{\vk}\dg \smvp e^{-i \vk \cdot \vp} }
.
\end{align}
It is therefore convenient to write the $\sU$-superoperator (see \cref{sec:formalism}) as an analogous sum, i.e., $\sU=\sum_{\vk,\vp}\sUkp{\vk}{\vp}$ with $\sUkp{\vk}{\vp} = -\sL_0\ginv \sL_1^{\vk \vp}$. Each $\sUkp{\vk}{\vp}$ is off-diagonal with respect to the unperturbed basis, so that $\sUkp{\vk}{\vp}=\llbracket \sUkp{\vk}{\vp} \rrbracket = \sUp_1^{\vk\vp}$ holds.

The rank-1 term in the resummation [\cref{eq:resummed}] is  given by
\begin{widetext}
\begin{align}
\label{eq:rsc_1}
\oket{\rsc{1}}&=
\frac{1}{\openone - \sloop}\sum_{\vk, \vp}\sUkp{\vk}{\vp} \oketb{\rsoc{0}}
= \sum_{\vk, \vp}  \sum_{\vec{s}}	\frac{1}{\openone - \sloop}\oket{\rscp{1}{\vk \vp\vec{s} }}
\omatb{\rscpl{1}{\vk \vp \vec{s} }}{\sUkp{\vk}{\vp}}{\rsoc{0}}.
\end{align}
Here, $\oketn{\rscp{1}{\vk \vp\vec{s}}}=\oketn{\evrr{\vk}{m}{n} \evrq{\vp}{\mu}} \bigotimes_{\vk'\neq\vk}\oketn{\evrs{\vk'}}
\bigotimes_{\vp'\neq \vp}\oketn{\evqs{\vp'}}$ is an interacting cluster involving the photon mode $\vk$ and qubit at position $\vp$ in state $\vec{s}=(m,n,\mu)$.
In a similar manner, we see that the rank-2 term
\begin{align}
\label{eq:rsc_2}
\oket{\rsc{2}}&=
\frac{1}{\openone - \sloop}\sum_{\vk,\vk',\vp,\vp'}\llbracket  \llbracket\sUkp{\vk}{\vp} \rrbracket\sUkp{\vk'}{\vp'} \rrbracket \oketb{\rsoc{0}} 
=
\sum_{\vk,\vk',\vp,\vp'}\sum_{[\vec{s}\vec{s'}]} \frac{1}{\openone - \sloop}\oket{\rscp{2}{\vk \vp\vec{s};\vk'\vp'\vec{s'} }}\!\!
\omatb{\rscpl{2}{\vk \vp\vec{s};\vk'\vp'\vec{s'} }}{\sUkp{\vk}{\vp}}{\rscp{1}{\vk'\vp'\vec{s'} }}\!
\omatb{\rscpl{1}{\vk'\vp'\vec{s'}}}
{\sUkp{\vk'}{\vp'}}{\rsoc{0}}
\end{align}
\end{widetext}
incorporates the resummation of a cluster $\oketn{\rscp{2}{\vk \vp \vec{s}, \vk'\vp'\vec{s'}}}$ composed of two photon modes and two qubits in states $\vec{s}$ and $\vec{s'}$. (The bracket notation in the corresponding summation signals that the allowed choices of these states is dictated by the off-diagonalism requirements in $\sUp_2=\llbracket  \llbracket\sUkp{\vk}{\vp} \rrbracket\sUkp{\vk'}{\vp'} \rrbracket $.) In the definition of $\oketn{\rscp{2}{\vk \vp \vec{s}, \vk'\vp'\vec{s'}}}$, several cases must be distinguished according to whether $\vk=\vk'$ and/or $\vp=\vp'$. For the case where both pairs are distinct, we have
\[
\oketn{\rscp{2}{\vk \vp \vec{s}, \vk'\vp'\vec{s'}}} =\oketn{\evrr{\vk}{m}{n}\evrr{\vk'}{m'}{n'}  \evrq{\vp}{\mu}\evrq{\vp'}{\mu'}} \bigotimes_{\vk''\neq\vk,\vk'}\oketn{\evrs{\vk''}}
\bigotimes_{\vp''\neq \vp,\vp'}\oketn{\evqs{\vp''}}.
\]
Analogous definitions hold in the other three cases.

By inspection of \cref{eq:rsc_1,eq:rsc_2} we expect that, in general, the rank-$j$ correction consists of a sum over all possible terms in which clusters of $j$ photon modes and $j$ qubits deviate from the unperturbed density matrix. Thanks to resummation, interaction within each cluster includes terms up to infinite order. We note that this cluster structure directly implies a hierarchy of correlations with increasing rank $j$. Specifically, every $n$-point correlation function with $p$ photon  and $q$ qubit operators, $\langle 
\bt_{\vk_1}^{(\dagger)}\cdots \bt_{\vk_p}^{(\dagger)}
\sigma^{a_1}_{\vp_1}\cdots \sigma^{a_q}_{\vp_q}\rangle_\text{ss}$, does not trivially separate into a product of correlators if the rank $j$ of the correction satisfies $j\ge \max \{p,q\}$. We also emphasize that clusters automatically include long-range correlations between distant qubits.

The essential tasks of determining the perturbative corrections and resummation consist of evaluating matrix elements of the form given in \cref{eq:rsc_1,eq:rsc_2}, and computing the effect of the resummation superoperator $\sloop$ to a given order.  We illustrate the procedure for the example of rank-1 corrections. Plugging in the definition of $\sUkp{\vk}{\vp}$ and recalling that $\sL_0$ is diagonal with respect to the unperturbed basis states, we obtain
\begin{align}
\omatb{\rscpl{1}{\vk \vp \vec{s} }}
{\sUkp{\vk}{\vp}}
{\evrs{\vk} \evqs{\vp}} &=
-
\omatb{\evlr{\vk}{m}{n} \evlq{\vp}{\mu}}
{\sL_1^{\vk \vp}}
{\evrs{\vk} \evqs{\vp}}/\lambda_{\vk\vp\vec{s}}\\\nonumber
&=i\Tr (\evlr{\vk\dag}{m}{n} \evlq{\vp\dag}{\mu}\,[\sfH_{rq}^{\vk \vp},\evrs{\vk} \evqs{\vp}])/\lambda_{\vk\vp\vec{s}}
\end{align}
where $\lambda_{\vk\vp\vec{s}} = 
\evarvk{m}{n} + \evaqvp{\mu}$. Once the commutator is opened, it is useful to note that the simple properties of the $\sL_r$-eigenstates lead to vanishing overlaps $\Tr (\evlr{\vk\dag}{m}{n}\bt_\vk^{(\dagger)}\evrr{\vk}{m}{n})=
\Tr (\evlr{\vk\dag}{m}{n}\evrr{\vk}{m}{n}\bt_\vk^{(\dagger)})=0$, so that any application of $\sUkp{\vk}{\vp}$ must switch to a different resonator-mode eigenstate. The same does not hold for traces of the qubit degrees of freedom, i.e., the overlaps $\Tr (\evlq{\vk\dag}{\mu}\sigma^+_\vp\evrq{\vk}{\mu})$ etc.\ may indeed be non-zero. As a result, we obtain the two types of terms for the rank-1 correction which are diagrammatically depicted in \cref{fig:diagram_representation}(a). The evaluation of rank-2 corrections follows the same basic scheme. Unsurprisingly, it is more tedious and we only show two examples of corresponding diagrams in \cref{fig:diagram_representation}(b). 

\begin{figure}
	\centering
	\includegraphics[width=0.95\columnwidth]{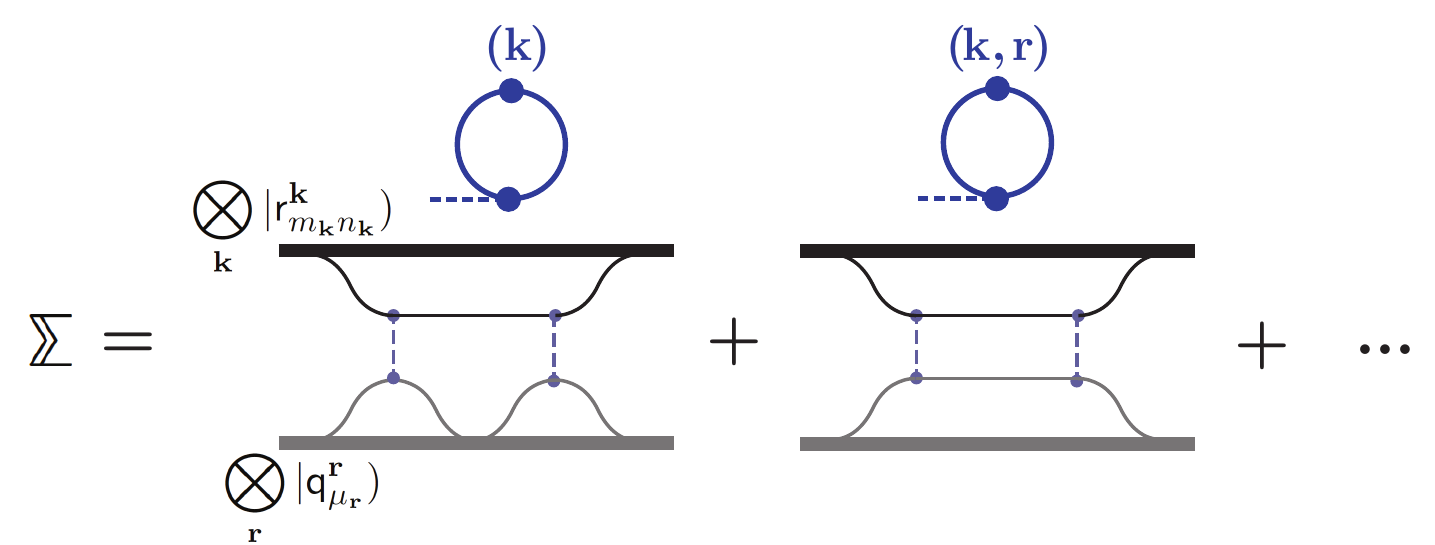}
	\caption{Evaluation of the $\sloop$-superoperator. $\sloop$ is needed for resummation and composed of irreducible diagrams starting end terminating in the same state. The two diagrams show the leading-order contributions, $\sloopn{2}$. 
	}
	\label{fig:sigmadiagram}
\end{figure}

The effect of the resummation superoperator is to redistribute weights among cluster contributions. Since $\sloop$ is diagonal with respect to unperturbed Liouvillian eigenstates, we can cast its contribution to a particular into the form 
$\sloop|\evrc{\vec{s}}{0})=|\evrc{\vec{s}}{0})(\evlc{\vec{s}}{0}|\sloop|\evrc{\vec{s}}{0})$ and evaluate the occurring matrix element as follows. We choose an appropriate truncation for the series $\sloop=\sloopn{2}+\sloopn{3}+\cdots$ of irreducible resummation operators $(\sloopn{j})_{\vec{s}\vec{s'}}=\delta_{\vec{s}\vec{s'}}(\sUpc{j-1}\sU)_{\vec{s}\vec{s'}}$. Matrix elements for $\sloopn{j}$ are  calculated in the same way as for $\oketn{\rsc{j}}$ except that the final state of the chain must be identical to the initial state. \cref{fig:sigmadiagram} shows the resulting two diagrams for $\sloopn{2}$.

If we are merely interested in certain steady-state expectation values (rather than in the density matrix itself), then the calculation of perturbative results may be simplified. As an example, consider computing an expectation value of a local qubit operator $\sigma_\vp^a$ up to corrections of rank $j$, $\langle \sigma_\vp^a \rangle \approx \sum_{j'=0}^j \Tr (\sigma_\vp^a \rsc{j'})$. To effect the desired simplification, we recall that all eigenstates of $\sLr{\vk}$ and $\sLq{\vp}$ other than the steady state must be traceless, i.e., $\Tr(\evrq{\vp}{\mu}) =0 $ for $\mu \neq 0$ and $  \Tr(\evrr{\vk}{m}{n}) = 0$ for non-zero $m$ or $n$ \footnote{To quickly confirm the tracelessness, refer to the orthonormality condition \eqref{eq:bi_orthonormal_condition} and note that the left eigenstate corresponding to the steady state is the identity, $(\rscl{0}|= \openone$.}. Therefore, any perturbative contribution $\sim\Tr(\sigma^a_\vp \evrc{\vec{m}\vec{n}\vec{\mu}}{0})$ in which $\mu_{\vp'}\not=0$ for some $\vp'\not=\vp$ will immediately vanish since the partial trace over the qubit at position $\vp'$ is zero. Similarly, any term with $(m_\vk,n_\vk)\not=(0,0)$ for some photon mode $\vk$ will vanish.
As a result, none of the rank-1 corrections [\cref{fig:diagram_representation}(a)] contribute to local qubit expectation values. Only those diagrams that terminate in a state labeled $(\vp)$ will yield a nonzero contribution to $\langle \sigma_\vp^a \rangle$. Analogous diagrammatic rules apply for photon-mode operators.

\subsection{Perturbative results for a near-resonant regime}
\label{sec:Result}
We now illustrate the validity and improvement achieved by (partial) resummation of the Lindblad perturbation series. For this purpose, we compare perturbative results for finite-size Jaynes-Cummings chains [\cref{fig:lattice_structure}(a)] to exact results computed via quantum trajectories methods. 
Perturbative resummation further allows us to treat periodic chains (or, open chains if desired) of sizes beyond the computational capabilities of exact quantum-trajectory solutions. Finally, we can carry out perturbative resummation even for an infinite system with chain or global-coupling geometry.
This versatility enables us to predict finite-size effects, the approach to the thermodynamic limit, as well as differences according to distinct lattice geometries. (We discuss the quite moderate computational costs of the perturbative treatment in \cref{app:ccost}.)

In our treatment, we capture photon mediated qubit-qubit interactions
by second-order Lindblad perturbation theory with resummation based on single-loop terms $\sloopn{2}$ (i.e., corrections of rank 2 in the above terminology). The most natural regime for treating the Jaynes-Cummings coupling perturbatively in this way is the dispersive regime where the detuning $\Delta = \min_\vk |\Omega - \omega_\vk|$ between   qubit and photon-mode frequencies is large compared to their mutual coupling strength $g$ \cite{Zhu2013b}. We have confirmed by exact numerics that the perturbation theory is indeed reliable in this regime and, over a wide range of parameters, we identify $g^2/\gaq \Delta$ as the relevant small parameter governing the series expansion.

In the following, we choose to present results from exploring a different parameter regime more directly based on the open-system nature of Jaynes-Cummings lattices. Specifically, we consider the case where photon hopping dominates over both photon decay and Jaynes-Cummings coupling, and where the latter two are permitted to be of the same order, i.e.\ $\ka \gg g \sim \gar$. The strong hopping $\ka$ shifts spectral weight of the photon modes away from the bare resonator frequency which is chosen degenerate with the qubit frequency, $\Omega=\omega$. This regime is not fully dispersive, so that nonlinearities are more pronounced, and the significance of resummation becomes easily visible.

\begin{figure*}
	\centering
	\includegraphics[width=1.0\textwidth]{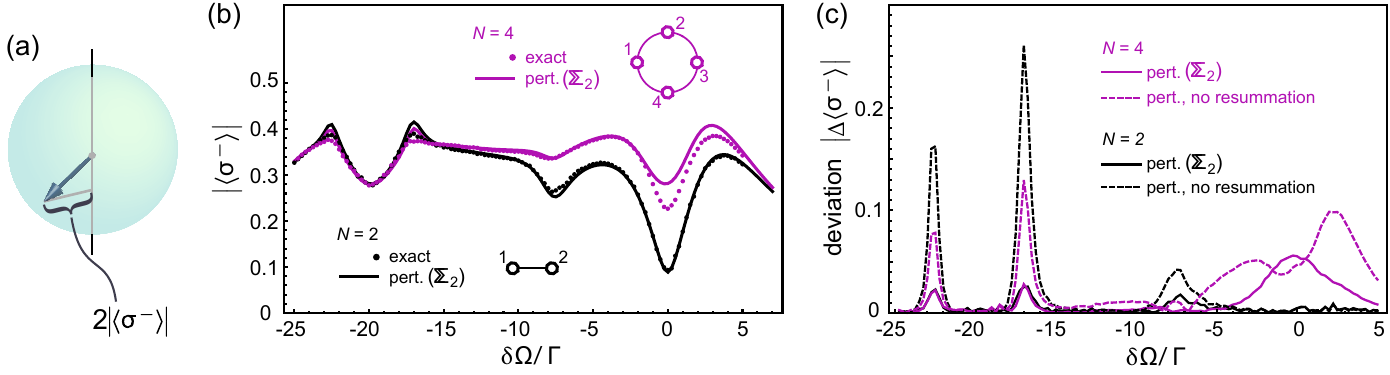}
	\caption{ Comparison between perturbative results and exact solution.
		\textbf{(a)} Within the Bloch-sphere picture, the steady-state expectation value $\abs{\savg{\smi{}}}$ is directly proportional to the distance of the Bloch vector (representing $\rs^\text{qubit}$) from the $z$ axis.
		\textbf{(b)} shows $\abs{\savg{\smi{}}}$ as a function of detuning $\dwq$ between drive and bare qubit frequency, in units of the qubit relaxation rate $\Gamma$. Qubits are held in resonance with the bare resonator frequency,   $\Omega = \omega$. Exact and perturbative results for chain sizes $N=2$ and $4$ are in good agreement. (See text for explanation of deviations close to $\dwq=0$ in the $N=4$ case.)
		\textbf{(c)} Curves here depict the absolute deviations between exact and perturbative results for calculations with (solid lines) and without resummation (dashed lines). Perturbation theory is not sufficient to describe the region close to $\dwq=0$ for $N=4$ (see text). Outside this region, resummation consistently improves agreement with the exact solution. [Parameters:  $g/ \gaq = 3$, $\ep /\gaq =20$, $\ka /\gaq =10$, and $\gar / \gaq = 4$.]}
	\label{fig:finite_size_result}
\end{figure*}

We begin with the comparison between perturbative and exact results for the steady state of few-site Jaynes-Cummings chains with periodic boundary conditions. In our calculations, we have considered several qubit and resonator expectation values. Among those, we find that $\abs{\savg{\smi{}}}=\sqrt{\langle \sigma^x\rangle^2 +\langle \sigma^y\rangle^2}/2$ is a convenient choice for clearly resolving resonances. Representing the reduced steady-state density matrix for one of the qubits by means of the Bloch sphere picture,  this quantity is directly proportional to the distance of the Bloch vector from the $z$ axis, see \cref{fig:finite_size_result}(a). Computing exact steady-state solutions for Jaynes-Cummings chains even as small as 4 sites, is a non-trivial task which we accomplish by averaging of quantum trajectories.  For instance,  exact results presented in \cref{fig:finite_size_result} for $N=4$ were determined from stochastic time evolution of a quantum state vector of size $10^4$. Sufficient averaging of a single data point required a runtime of several days on one core. 

\begin{figure*}
	\centering
	\includegraphics[width=0.85\textwidth]{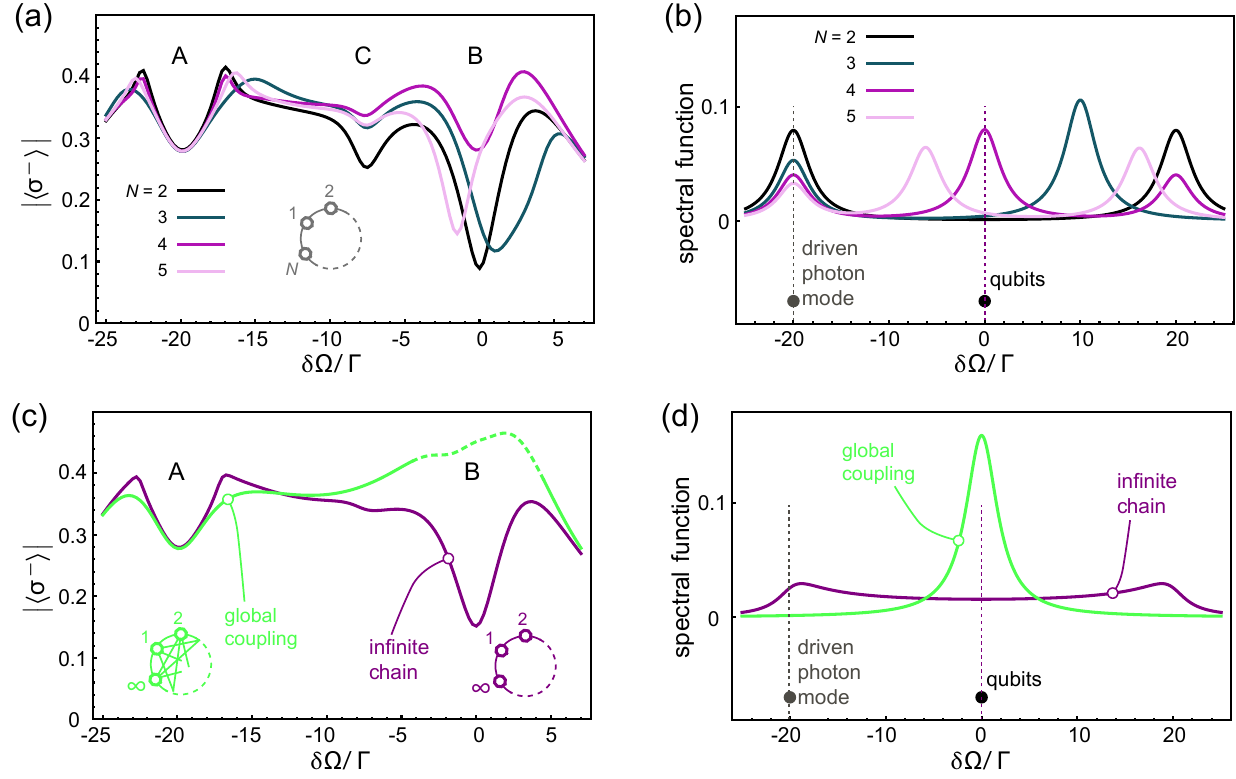}
	\caption{Perturbative results and spectral functions for  periodic Jaynes-Cummings chains and global-coupling model.  \textbf{(a)} shows the qubit expectation value $\abs{\savg{\smi{}}}$ for periodic chains with different site numbers $N$. We observe three resonance dips (labeled A--C). A and B are located roughly at the driven photon mode and qubit frequencies, respectively.
	\textbf{(b)} Spectral functions resonator modes differ for small site numbers. Presence and position of resonances near the qubit frequency explain the $N$-dependence of strength and shift of the B resonance.
	\textbf{(c,d)} show analogous plots for infinite lattices with periodic chain or global-coupling geometry. Resonances A and B are visible, but resonance C is (nearly) absent.
	[All system parameters are identical to those in  \cref{fig:finite_size_result}.]
	}
	\label{fig:PT_result}
\end{figure*}

The comparison between exact and rank-2 perturbative data (including resummation at the level of $\sloopn{2}$) in \cref{fig:finite_size_result}(b) shows very good agreement, and indicates that the resummation procedure closely matches the exact solution. Plots in this figure show the steady-state value of $\abs{\savg{\smi{}}}$ as a function of the detuning $\dwq=\Omega-\omega_d$ between the drive and the bare qubit frequency. Multiple resonances are visible over the chosen frequency range (the nature of which we will further discuss below). The only notable quantitative deviations occur for $N=4$ in the vicinity of the bare qubit frequency where $\dwq=0$. This deviation has a simple explanation: a look ahead at \cref{fig:PT_result}(b) shows that the 4-site chain has a photon mode with large spectral weight directly on resonance with the bare qubit frequency, so that there we must expect the perturbative treatment in $g$ to break down. 
With the exception of this finding,  we conclude that resummation of the perturbative series in the chosen parameter regime works very well. The improvement gained over the pure second-order approximation is illustrated in \cref{fig:finite_size_result}(c). In this panel, curves show the difference between approximate and exact results, $\Delta\savg{\smi{}} =\savg{\smi{}}_\text{approx} - \savg{\smi{}}_\text{exact} $, for the case  including resummation (solid lines) and lacking resummation (dashed lines). Excluding the pathological region for $N=4$ around $\dwq=0$, we observe that resummation consistently improves the results, reducing the deviation from the exact solution. The improvement is especially significant in the resonance region between $\dwq/\gaq \approx -17$ and $-23$. Here, the drive populates the uniform mode (centered at $\dwq/\gaq=-20$) and renders photon-mediated qubit-qubit interaction important, making resummation of corrections up to infinite order particularly fruitful.

We next turn to the discussion of the resonances visible in \cref{fig:finite_size_result} and shown for additional site numbers $N$ in \cref{fig:PT_result}(a).  For small chain lengths up to $N=5$ sites,  we observe three resonances labeled $\mathsf{A}$, $\mathsf{B}$, and $\mathsf{C}$ in the range of drive frequencies spanning the photon-chain eigenmodes and qubit frequencies. [The uniform photon mode has the lowest frequency, $\dwq/\gaq = -20 $, and the bare qubit frequency is at $\dwq\gaq = 0 $, see vertical lines in \cref{fig:PT_result}(b)].
We find that the detailed positions and strengths of resonances depends  on the number of sites, revealing systematic finite-size effects for chains of short lengths. Both the nature and $N$ dependence of resonances can be explained, or at least motivated, by the following considerations.

The resonance marked $\mathsf{A}$ directly coincides with the frequency of the uniform photon mode. Equivalent interpretations of the resonance can be given based on the original Hamiltonian \cref{eq:JCH} with a coherent tone driving this particular mode with strength $\ep$, or for the Hamiltonian following the displacement transformation \cref{eq:displaced_H,eq:displaced_qubit}. Employing the language of the latter description, we note that the strength $\epe=-g\ep/(\dwrm{\vk=\vm}-i\gamma/2)$ of the effective qubit drive reaches its local maximum at the uniform-mode frequency ($\dwq/\gaq = -20 $). This peak in the off-resonant Rabi drive, modified by weak coupling to photon modes, is responsible for resonance $\mathsf{A}$. Its dependence on the site number $N$ is relatively weak and mainly affects the shoulders of the resonance. This is further confirmed by our results for the infinite-system case with periodic-chain and global-coupling geometry [\cref{fig:finite_size_result}(c)]. Both show the same resonance $\mathsf{A}$, but differ in the resonance shoulders.

For resonance $\mathsf{B}$, the $N$ dependence of resonance position and strength is much more pronounced. The general region where $\mathsf{B}$ occurs is close to $\dwq=0$, i.e., where bare qubit frequency and drive frequency match. Upon displacement, the drive transforms into an effective Rabi drive on each qubit [\cref{eq:displaced_qubit}]. Hence, the presence of a resonance is natural, and variations in its strength and precise position must be governed by the Jaynes-Cummings interaction playing the role of the perturbation in our treatment [\cref{eq:JC_perturbation}]. The importance of this interaction is influenced by the position of photon-mode resonances $\omega_\vk$ relative to the bare qubit frequency. In \cref{fig:PT_result}(b) and (d), we depict width and position of photon modes in terms of the spectral function 
 $s(\omega)=\sum_{\vk}\tfrac{\gar/2}{(\omega-\omega_\vk)^2 + (\gar/2)^2}$, which is the sum over individual Lorentzians for each photon mode, normalized such that $\int d\omega\, s (\omega)= 1$.
Inspection of resonance-$\mathsf{B}$ positions [\cref{fig:PT_result}(a),(c)] and peaks in the spectral function [\cref{fig:PT_result}(b),(d)] shows that peaks in $s(\omega)$ with significant weight in the region $\Omega\pm g$, shift $\mathsf{B}$ resonances towards the close-by photon mode (such as for $N=3,5$).  Further, it is clear that strongly increased weight of the spectral function directly at the qubit frequency (such as for $N=4$ and for the global-coupling geometry) endangers the validity of perturbation theory in the Jaynes-Cummings coupling. Above, we recognized this as the reason for the observed deviations between perturbation theory and exact solution close to $\delta\Omega=0$ in the $N=4$ case. A look at the spectral function for the global-coupling geometry shows that the same issue occurs here. Accordingly, we show the perturbative result in \cref{fig:PT_result}(c) only with dashes in that region.

We note that steady-state expectation values for infinite lattices are not always easily accessible by other methods. Thanks to the possibility of carrying out leading-order resummation  analytically in the infinite-system case, our treatment gives direct access to the thermodynamic limit of different lattice geometries. Here, we have chosen two extreme cases: the infinite periodic chain with a minimum number of links between sites, and the global-coupling model with a maximum number of links. 
\cref{fig:PT_result}(c) depicts results for both lattice structures. We expect that the region close to $\dwq=0$ is unproblematic for the infinite chain case, but potentially  pathological for the global-coupling model which accumulates maximum spectral weight at the bare qubit frequency [\cref{fig:PT_result}(d)]. Away from the $\dwq=0$ range, the two geometries yield similar behavior of $\abs{\savg{\smi{}}}$ versus drive detuning $\delta\Omega$. As before, characteristic differences occur primarily in the shoulders of resonance $\mathsf{A}$. Interestingly, the resonance $\mathsf{C}$ is absent for both infinite lattices, and we discuss the rather anomalous behavior of this finite-size resonance next.

\begin{figure*}
	\centering
	\includegraphics[width=0.85\textwidth]{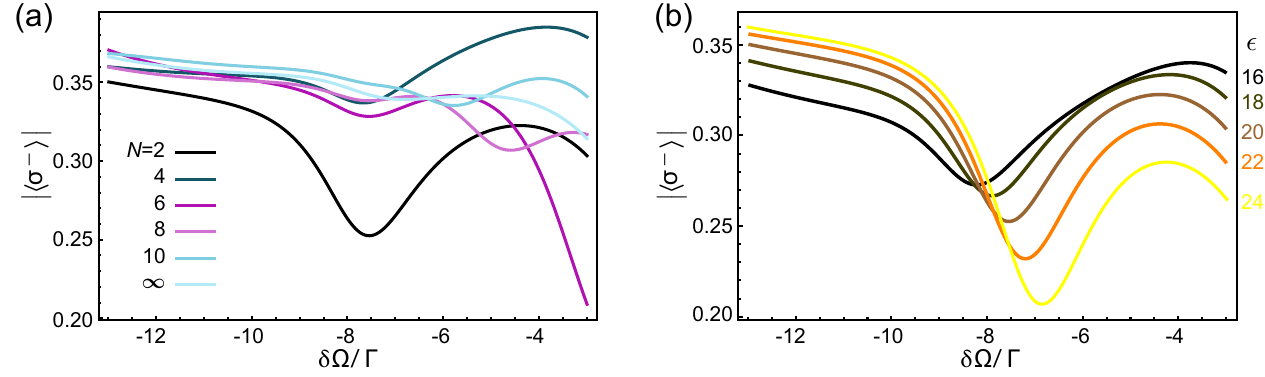}
	\caption{Dependence of the anomalous resonance C on site number $N$ and drive strength $\ep$. 
		\textbf{(a)} The anomalous dip becomes less prominent with increasing $N$ (even) and nearly vanishes for the infinite chain. (The same trend applies to odd site numbers.)
		\textbf{(b)} For the dimer case, $N=2$, the position of the anomalous resonance shifts monotonically with increasing drive strength $\ep$. The same trend is observed for longer chains. [Parameters are chosen the same as in \cref{fig:finite_size_result}.]}
	\label{fig:PT_dip}
\end{figure*}

The anomalous properties of resonance $\mathsf{C}$ are illustrated in \cref{fig:PT_dip}. The position of this resonance close to $\dwq/\Gamma\approx 8$ does not simply coincide with a resonance between photon modes and bare qubits. Panel (a) shows the decrease of the resonance strength for increasing chain length $N$. Moreover, both resonance position and strength depend sensitively on the drive power $\sim\ep$ as shown in \cref{fig:PT_dip} for the dimer case, $N=2$. We investigate this anomaly for the $N=2$ case, where a semi-quantitative reduced model can shed light on the origin and nature of this resonance. 

\begin{figure}
\centering
\includegraphics[width=\columnwidth]{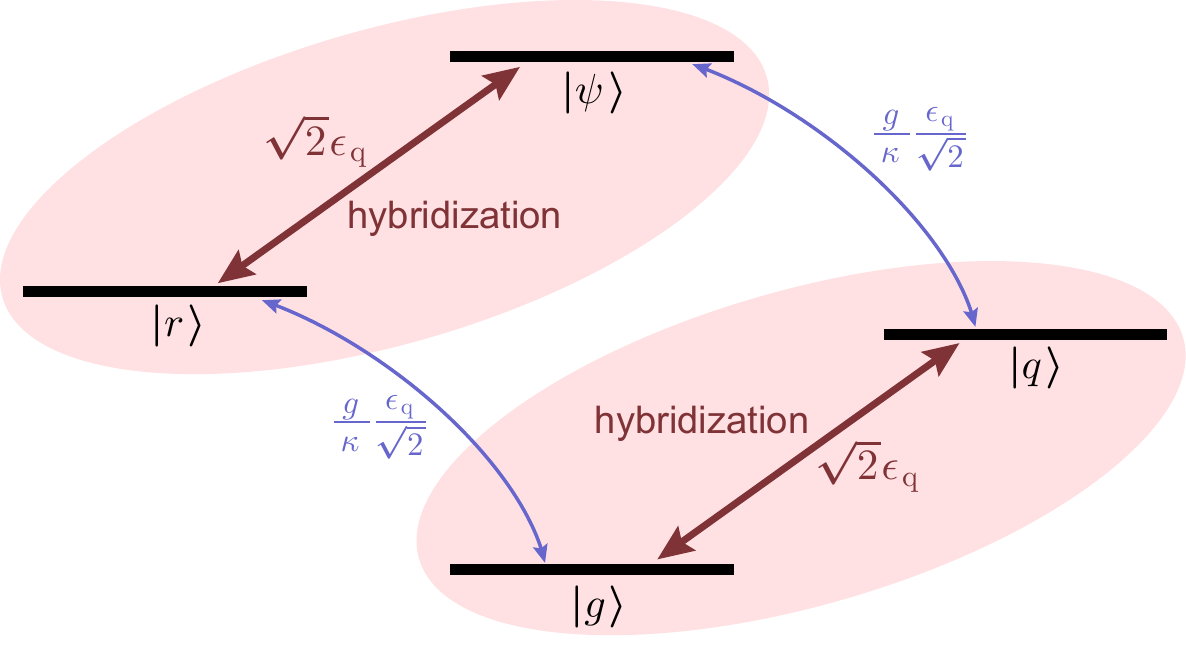}
\caption{Effective four-level model explaining the anomalous resonance. The transitions $\lf| g \rt> \leftrightarrow \lf| q \rt>$ and $\lf| r \rt> \leftrightarrow \lf| \psi \rt>$ strongly hybridize two pairs of states. The anomalous resonance results from transitions between the two hybridized doublets. Since the energy separation between hybridized doublets depends on drive strength $\ep$, so does the position of the anomalous resonance.}
\label{fig:effective_model}
\end{figure}

For $N=2$ we can confirm analytically that the anomalous resonance $\mathsf{C}$ is closely related to an eigenstates $\lf| \psi \rt>$ of the displaced Hamiltonian $H'$ [\cref{eq:displaced_H}] without the effective drive. This eigenstate comprises two excitations distributed between the uniform mode and the qubits. Truncation to first order in $g/\ka$ (a small quantity for the chosen parameter set) yields
\begin{align}
\lf| \psi \rt> \approx & 
\frac{1}{\sqrt{2}} \lf|1_{ \vm}\rt>  (\lf| \uparrow \downarrow \rt> + \lf| \downarrow \uparrow \rt>)
\nn\\
&
- \frac{1}{\sqrt{2}} \frac{g}{\ka} \lf|2_{ \vm}\rt> \lf| \downarrow \downarrow \rt>  + \frac{1}{2} \frac{g}{\ka} \lf| 0_{ \vm} \rt> \lf| \uparrow \uparrow  \rt>
\label{eq:psi}
\end{align}
with eigenenergy $2 \Omega + 2 \kappa - g^2/2\kappa$.
Here, $\lf| n_{ \vm} \rt>$ is the $n$-photon state of the uniform mode and $\lf| \uparrow  \uparrow  \rt>$ etc.\ denote states of the qubits on the two sites.
The effective drive Hamiltonian with strength $\epe$ connects the ground state $\lf| g \rt> = \lf|0_{ \vm}\rt>  \lf| \downarrow \downarrow \rt> $ to the state $\lf| \psi \rt>$ via two intermediate states $\lf|r \rt>$ and $\lf| q \rt>$, see \cref{fig:effective_model}. These intermediate states belong to the one-excitation mani\-fold and primarily consist either of a photon in the uniform mode or or of a qubit excitation, respectively. Truncated again to first order in $g/\ka$, $\lf|r \rt>$ and $\lf| q \rt>$ are given by
\begin{align}
\lf| r \rt> \approx &
\lf|1_{ \vm}\rt>\lf| \downarrow \downarrow \rt>   + \frac{1}{2\sqrt{2}} \frac{g}{\ka} \lf| 0_{ \vm} \rt>  (\lf| \uparrow \downarrow \rt> + \lf| \downarrow \uparrow \rt>),
\\
\lf| q \rt> \approx &
\frac{1}{\sqrt{2}} \lf| 0_{ \vm} \rt>  (\lf| \uparrow \downarrow \rt> +   \lf| \downarrow \uparrow \rt>) - \frac{1}{2} \frac{g}{\ka} \lf|1_{ \vm}\rt>\lf| \downarrow \downarrow \rt>  .
\end{align}
Our description of the anomalous resonance  in the following is based on the effective four-level model spanned by the states $\lf| g \rt>$, $\lf| r \rt>$, $\lf| q \rt>$, and $\lf| \psi \rt>$, see \cref{fig:effective_model}.

Within this model, the effective drive Hamiltonian connects the ground state $\lf| g \rt>$ to the two-excitation state $\lf| \psi \rt>$ via $\lf| r \rt>$ and $\lf| q \rt>$. Since the effective drive creates (or annihilates) qubit excitations only, there is  stronger hybridization of $\lf| g \rt>$ with $\lf| q \rt>$ (strength $\sim\epe$) than with $\lf| r \rt>$ (strength $\sim \epe\,g/\kappa$). An analogous argument applies to explain hybridization of $\lf| r \rt>$ with $\lf| \psi \rt>$ (again, strength $\sim\epe$)). We thus obtain the picture of two pairs of hybridized states, $\lf| g \rt> \leftrightarrow \lf| q \rt>$ and $\lf| r \rt> \leftrightarrow \lf| \psi \rt>$, with only small drive matrix elements connecting the pairs.

Due to the energy differences between states within each pair, hybridization is only partial. The two emerging hybridized states relevant to the anomalous resonance have significant overlap with the ground state $\lf| g \rt>$ in one case, and the two-excitation state $\lf| \psi \rt>$ in the other case. The resonance $\mathsf{C}$ can be approximately viewed as a resonance between these two hybridized states. Note that the degree of hybridization critically depends on the effective drive strength $\epe$, which is in turn proportional to the drive strength $\ep$. As a consequence, the energy separation between the two relevant hybridized states depends on drive strength as observed in \cref{fig:PT_dip}(b).

The generalization of the effective four-level model to periodic chains with larger number of sites $N$ is difficult due to the proliferation of degeneracies among eigenstates of $H'$ in the absence of a drive. Based on our perturbative calculations, we find a clear trend of diminishing resonance strength with increasing number of sites [\cref{fig:PT_dip}(a)]. The anomalous resonance $\mathsf{C}$ hence provides an interesting example of an interaction-induced feature which is limited to a small $N$, which should be accessible in experiments with Jaynes-Cummings chains of only a few sites. 

\section{Conclusion}
\label{sec:conclusion}
We have extended the general Lindblad perturbation framework to include  resummation of an infinite subset of perturbative corrections. We have formulated the scheme at a general level, emphasizing that is not limited to a particular open quantum system, but benefits a variety of problems in Markovian quantum systems amenable to perturbative treatment.  For the examples we have investigated, we find that the series resummation can significantly improve the accuracy of the perturbative treatment.

We have applied perturbation theory with resummation to a specific model of an open quantum lattice, the open Jaynes-Cummings lattice, and have introduced a diagrammatic representation systematically organizing the contributing terms.  For small lattices, we find very good agreement with exact results which we obtained by extensive quantum trajectory simulations for an interesting parameter regime near resonance.
Our perturbative treatment is capable of predicting steady-state observables  for both finite and infinite Jaynes-Cummings lattices with different lattice geometries and dimensionalities -- thus including settings that may not be easily accessible by other methods. 
 
The capability of obtaining reliable results beyond exactly solvable limits of open quantum lattices is particularly promising as a method for validating experimental implementations of quantum simulators. Concrete realizations of open quantum lattices are currently being investigated in the circuit QED architecture \cite{Houck_inprep}.

Finally, questions that warrant future investigations regard the relation between mean-field approximations and suitable choices of resummation within the presented formalism. While Keldysh techniques tend to be more difficult for handling (pseudo-)spin degrees of freedom, it will be interesting to compare  and relate results obtainable for simpler systems such as the open Bose-Hubbard lattice. Further investigations into spectra of Liouvillians, handling cases of degeneracies of the Liouvillian spectrum, and extending resummation to time-dependent perturbation theory offer exciting perspectives for the study of dissipative phase transitions in open quantum systems.

\begin{acknowledgments}
We acknowledge valuable discussions with Guanyu Zhu and Joshua Dempster, and thank Ata\c{c} \.Imamo\u{g}lu, Juan Jos\'e Garc\'ia-Ripoll, and Mark I.\ Dykman for pointing out useful references developing and applying Lindblad perturbation theory.  This research was supported in part through the computational resources and staff contributions provided for the Quest high performance computing facility at Northwestern University, by the NSF under Grant No.\ PHY-1055993 (A.C.Y.L. and J.K.), and by the South African Research Chair Initiative of the Department of Science and Technology and National Research Foundation (F.P.). J.K. thanks KITP (Santa Barbara) for hospitality and financial support through NSF Grant No.\ PHY11-25915.
\end{acknowledgments}

\appendix
\section{Details of resummation} \label[appendix]{app:proof_SE}
This appendix provides the proof that powers of the $\sU$-superoperator can be written as [\cref{eq:pattern}],
\be
\label{eq:app_pattern}
\sU^j = \sum_{k=0}^{j} \spSE{j-k} \sUpc{k},
\ee
where the prefactor $\spSE{j-k}$ consists of all possible combinations of $\sloopn{}$-superoperators of total order $j-k$.
The definitions of the involved $\sUpc{}$-superoperator and $\sloopn{}$-superoperator [\cref{eq:def_SE_0,eq:separation_SE}] are
\begin{align}
\label{eq:app_def_SE_0}
&\sloopn{1} = 0 , \qquad
\sUpc{1} = \sU,
\\
\label{eq:app_separation_SE}
&\sUpc{j-1} \sU = \sloopn{j}+ \sUpc{j}, \qquad
\lf(\sloopn{j}\rt)_{\mu \nu}   = \delta_{\mu \nu} \lf( \sUpc{j-1} \sU \rt)_{\mu \nu}.
\end{align}
We further define $\spSE{0}= \sUpc{0} = \openone$.

We prove \eqref{eq:app_pattern} by mathematical induction. The $\sU^1$ case clearly satisfies \eqref{eq:app_pattern},
\be
\sU^1 = \sUpc{1} = \sloopn{1} \openone + \openone \sUpc{1} = \spSE{1} \sUpc{0} + \spSE{0} \sUpc{1} .
\ee
Assume that \eqref{eq:app_pattern} holds up to power $j-1$, i.e.\
\be
\sU^{j-1} = \sum_{k=0}^{j-1}\spSE{k} \sUpc{j-1-k}.
\ee
Then, the decomposition rule \eqref{eq:sU_decomposition_rule} yields
\be
\sU^{j} = \sU^{j-1} \sU = \sum_{k=0}^{j-1} \spSE{k} \sUpc{j-1-k} \sU.
\ee
The product $\sUpc{j-1-k} \sU$ is separated according to \cref{eq:app_separation_SE}, i.e.\ $\sUpc{j-1-k} \sU = \sloopn{j-k}+ \sUpc{j -k}$, so we obtain
\be
\sU^{j}  = \sum_{k=0}^{j-1} \spSE{k} \sloopn{j-k}+ \sum_{k=0}^{j-1} \spSE{k} \sUpc{j -k}.
\ee
The first sum, $ \sum_{k=0}^{j-1} \spSE{k} \sloopn{j-k} $, consists of \emph{all} products of $\sloop$-superoperators with combined order $j$,
 i.e.\ $\sum_{k=0}^{j-1} \spSE{k} \sloopn{j-k} = \spSE{j}$. As a result, it follows that
\be
\sU^{j}  = \spSE{j}\sUpc{0}+ \sum_{k=0}^{j-1} \spSE{k} \sUpc{j -k} = \sum_{k=0}^{j} \spSE{k} \sUpc{j -k},
\ee
 which concludes the proof.

\section{Computational cost for perturbative calculations} \label[appendix]{app:ccost}
The computational cost in calculating  perturbative results for the open Jaynes-Cummings lattice primarily stems from summation over matrix elements of the $\sUkp{\vk}{\vp}$-superoperator. For each summation, the number of terms is given by $N$, which corresponds to either the number of qubits or photon modes. The number of necessary summations for rank-$j$ corrections is given by $2j$, see \cref{eq:rsc_1,eq:rsc_2}. As a result, the overall cost  scales algebraically with the number of sites, namely $\sim N^{2j}$. 
Results for infinite lattices hinge upon the possibility to carry out summations analytically for specific cases, such as for leading-rank corrections of infinite Jaynes-Cummings lattices with periodic-chain or global-coupling geometry.

\end{document}